\documentclass[a4paper,review]{cas-sc}

\usepackage[section]{placeins} 
\usepackage{amssymb}
\usepackage{amsthm}
\usepackage{amsmath}
\usepackage{upgreek}
\usepackage{pdflscape}
\usepackage{listings}
\usepackage{multirow}
\usepackage[percent]{overpic}
\usepackage{multicol}
\usepackage{color}
\usepackage{stmaryrd}
\usepackage{xfrac}
\usepackage[capitalise]{cleveref}
\usepackage{booktabs}
\usepackage[section]{placeins} 
\usepackage[section]{algorithm}
\usepackage{calc}
\usepackage[titletoc]{appendix}
\usepackage{siunitx}
\usepackage[sort&compress,square,numbers]{natbib}

\usepackage{graphicx}
\usepackage{graphics}
\usepackage{wrapfig}
\usepackage{float}
\usepackage{subfig}
\usepackage[percent]{overpic}
\usepackage{varwidth}
\usepackage{tikz}
\usepackage{pgfplots}
\usepackage{adjustbox}
\usetikzlibrary{arrows,matrix,positioning,fit}
\usetikzlibrary{shapes,positioning}
\usetikzlibrary{backgrounds}
\usepackage{tikz-layers}
\usepgfplotslibrary{fillbetween}
\usetikzlibrary{intersections}
\pgfplotsset{compat=1.14}
\usepgfplotslibrary{colorbrewer}
\usepgfplotslibrary{patchplots}
\usepgfplotslibrary[colorbrewer]
\usetikzlibrary{pgfplots.colorbrewer}
\usetikzlibrary[pgfplots.colorbrewer]
\usepgfplotslibrary{units}
\usetikzlibrary{spy}
\usepackage{pgfplotstable}
\usepackage{arrayjobx}
\graphicspath{ {./figs/} }
\usetikzlibrary{external}

\newlength\myheight
\newlength\mydepth
\settototalheight\myheight{Xygp}
\newcommand*\inlinegraphics[1]{%
  \settototalheight\myheight{Xygp}%
  \settodepth\mydepth{Xygp}%
  \raisebox{-\mydepth}{\includegraphics[height=\myheight]{#1}}%
}
\newcommand\orcid[1]{\href{https://orcid.org/#1}{\inlinegraphics{orcid_16x16.png}}}

\makeatletter
\def\BState{\State\hskip-\ALG@thistlm}
\makeatother

\newdefinition{definition}{Definition}[section]

\newcommand{\etal}{et~al.}

\begin{document}

\title[mode=title]{Partially-Averaged Navier-Stokes Simulations of Turbulence Within a High-Order Flux Reconstruction Framework}
\shorttitle{PANS Simulations of Turbulence Within a High-Order FR Framework}
\shortauthors{T. Dzanic~\etal}

\author[1]{T. Dzanic}[orcid=0000-0003-3791-1134]
\cormark[1]
\cortext[cor1]{Corresponding author}
\ead{tdzanic@tamu.edu}
\author[1]{S. S. Girimaji}[orcid=0000-0001-9443-435X]
\author[1]{F. D. Witherden}[orcid=0000-0003-2343-412X]

\address[1]{Department of Ocean Engineering, Texas A\&M University, College Station, TX 77843}

\begin{abstract}
High-order methods and hybrid turbulence models have independently shown promise as means of decreasing the computational cost of scale-resolving simulations. 
The objective of this work is to develop the combination of these methods and analyze the effects of high-order discretizations on hybrid turbulence models, particularly with respect the optimal model parameters and the relative accuracy benefits compared to approaches such as under-resolved direct numerical simulation (URDNS). 
We employ the Partially-Averaged Navier-Stokes (PANS) approach using the flux reconstruction scheme on the flow around a periodic hill and the wake flow of a circular cylinder at a Reynolds number of 3900, the latter of which we provide direct numerical simulation results and novel statistical analysis. By increasing the order of the discretization while fixing the total degrees of freedom, it was observed that larger improvements in the prediction of the statistics and flow physics were generally seen with PANS than URDNS. Furthermore, less sensitivity to the resolution-control parameter was observed with a high-order discretization, indicating that high-order discretizations may be an effective approach for increasing the accuracy and reliability of hybrid turbulence models for scale-resolving simulations without a significant increase in computational effort. 

\end{abstract}






\maketitle

\section{Introduction}
The efficient computation of complex turbulent flows remains a driving force in the development of computational fluid dynamics techniques. For a variety of practical engineering flows, the spatio-temporal resolution requirements of the underlying physical phenomena make direct numerical simulation (DNS) and large eddy simulation (LES) prohibitively expensive. Progress in this regard has generally followed two distinct paths: algorithmic advances and the development of higher fidelity subgrid-scale models. While the former approach attempts to reduce the computational cost, the goal of the latter is to reduce the resolution requirements without a significant detriment in accuracy. 

In the context of algorithmic design, a class of methods that have seen increased usage for high-fidelity simulations of turbulence over the past several decades are spectral element methods (SEM). These high-order methods offer the geometric flexibility of finite volume approaches without sacrificing the arbitrarily high order of accuracy that finite difference approaches can provide. Furthermore, due to their compact structure, they are well suited for modern massively parallel computer architectures, and when paired with the benefits of high-order accuracy such as the reduction in numerical dissipation, can significantly decrease the computational cost requirements of turbulent flows. Various spectral element methods have been applied to a wide variety of flows, with approaches such as the discontinuous Galerkin (DG) \citep{Reed1973, Hesthaven2008}, flux reconstruction (FR) \citep{Huynh2007}, and spectral difference (SD) \citep{Kopriva1996, Liu2004} methods being commonplace. However, their application has generally focused on LES and DNS \citep{Fernandez2017}, and as such, they suffer from the spatio-temporal resolution requirements of these techniques which prohibits their use for higher Reynolds number flows. 

The computationally unfeasible resolution requirements of complex flows have driven the development of higher fidelity subgrid-scale models. There exist a large variety of options in terms of techniques for simulating fluid flows. On one end of the spectrum, approaches such as Reynolds-Averaged Navier-Stokes (RANS) attempt to model all of the spatio-temporal scales of the flow, doing so with relatively little computational effort. However, the effectiveness of the approach is highly dependent on the problem and the model in question. On the other end, DNS attempts to resolve all of the spatio-temporal scales of the flow at the expense of tremendous computational cost. This cost is slightly alleviated through approaches such as LES in which only the statistically significant scales are resolved, either via an explicit approach using a filter with a subgrid-scale model or via an implicit approach using the numerical dissipation of the scheme, the latter of which is typically denoted as implicit LES (ILES) or under-resolved DNS (URDNS). However, the cost of these approaches generally makes them impractical for engineering applications, and as such, there is a necessity for methods that can offer higher fidelity than RANS at a lower cost than LES. These methods are typically denoted as scale-resolving simulation (SRS) techniques, which attempt to relax the resolution requirements of LES without sacrificing its ability to accurately resolve the predominant flow physics. One such class of techniques that have shown promise in this regard are hybrid turbulence models \citep{Girimaji2005,Schiestel2004,Speziale1998,Fasel2002,Pereira2018}. The intent of these SRS approaches is generally to resolve only the coherent flow structures while modeling the stochastic portion of the flow. As a result, their computational cost tends to scale much more reasonably with respect to flow complexity while potentially retaining many of the benefits of approaches such as LES in terms of accuracy. 

The unification of high-order numerical methods and hybrid turbulence models provides a potential for significant improvements in the computational cost of SRS.
Aside from the increased fidelity provided by high-order methods, the decrease in numerical dissipation can particularly benefit SRS methods as the excessive dissipation of low-order schemes has a much more detrimental impact on the modeled physics in comparison to RANS-type approaches. The application of high-order numerical methods to various turbulence models has been explored in the literature \citep{Ilinca1999, Bassi2005, Wurst2014, Nguyen2007}, with the majority of these works using high-order schemes for RANS or zonal methods such as detached eddy simulation (DES). However, the analysis on the actual effects of the high-order discretization on these turbulence models is extremely limited, particularly for SRS with hybrid turbulence models in which the numerical dissipation plays a more significant role and the accuracy of the discretization affects the resolution, making this analysis more complex. 

The goal of this work is therefore to analyze the effects of high-order discretizations on a hybrid turbulence model through a single numerical framework that can recover an arbitrary order of accuracy. For this, we employ the Partially-Averaged Navier-Stokes (PANS) approach of \citet{Girimaji2005}, a bridging turbulence model, using the FR approach of \citet{Huynh2007}, a discontinuous spectral element method. We study the effects of discretization order on the optimal resolution-control parameters, the relative accuracy benefits compared to approaches without physical models, and how the discretization affects the ability of the model to predict the dominant flow physics and flow structures. This analysis is performed on a wall-bounded, separated flow as well as the wake flow around a circular cylinder at a Reynolds number of 3900, the latter of which we provide DNS results and novel statistical analysis. The remainder of this paper is structured as follows. In \cref{sec:methodology}, the Partially-Averaged Navier-Stokes and flux reconstruction methods are presented, as well as implementation details and modifications to make the model more amenable to high-order discretizations. The numerical experiments are described in \cref{sec:numerical} followed by the results. Conclusions are then drawn in \cref{sec:conclusion}. 
\section{Methodology}\label{sec:methodology}

\subsection{Partially-Averaged Navier-Stokes Formulation}
The Partially-Averaged Navier-Stokes approach is derived from the application of a filter of arbitrary width to the Navier-Stokes equations \citep{Girimaji2005}. For the compressible form, this yields second moment and source terms in the resolved momentum and energy equations, which allow the governing PANS equations for the resolved density, momentum, and energy to be written as
\begin{align}
    \frac{\partial \rho}{\partial t} + \frac{\partial}{\partial x_i}(\rho u_i) &= 0,\\
    \frac{\partial}{\partial t}(\rho u_i) + \frac{\partial}{\partial x_j}(\rho u_i u_j) &= -\frac{\partial P}{\partial x_i} + \frac{\partial}{\partial x_j}\bigg [\tau_{ij} + \tau_{ij}'\bigg ],\\
    \frac{\partial}{\partial t}(\rho E) + \frac{\partial}{\partial x_j}(\rho E u_j) &= -\frac{\partial }{\partial x_j}(P u_j) + \frac{\partial}{\partial x_j}\bigg [u_i (\tau_{ij} + \tau_{ij}') - q_j + r_j'\bigg],
\end{align}
where $\rho$ is the density, $u_i$ are the velocity components, $\rho E$ is the total energy, $\gamma = 1.4$ is the ratio of specific heats, and $P = (\gamma-1) (\rho E - \frac{1}{2}\rho u_i u_i)$ is the pressure. The shear stress tensor $\tau_{ij}$ is taken as 
\begin{equation}
    \tau_{ij} = 2 \mu \bigg [ S_{ij} - \frac{1}{3} \frac{\partial u_k}{\partial x_k} \bigg ],
\end{equation}
where $\mu$ is the dynamic viscosity and 
\begin{equation}
    S_{ij} = \frac{1}{2} \bigg ( \frac{\partial u_i}{\partial x_j} + \frac{\partial u_j}{\partial x_i} \bigg ).
\end{equation}
Furthermore, the heat flux vector $q_j$ can be expressed as
\begin{equation}
    q_j = \frac{\mu}{Pr} \frac{\partial h}{\partial x_j},
\end{equation}
where $h$ is the specific enthalpy and $Pr = 0.71$ is the molecular Prandtl number.

To close the PANS equations, a constitutive relation for the additional terms, $\tau_{ij}'$ and $r_j'$, is required. By invoking the Boussinesq approximation and enforcing the conservation of unresolved turbulent kinetic energy, these terms can be written as
\begin{equation}
    \tau_{ij}' = 2\mu_u S_{ij} - \frac{2}{3} k_u \delta_{ij},
\end{equation}
\begin{equation}
    r_{j}' = -\frac{\mu_u}{Pr_t} \frac{\partial h}{\partial x_j} + \frac{\partial k_u}{\partial x_j},
\end{equation}
where $\mu_u$ is the unresolved eddy viscosity, $k_u$ is the unresolved turbulent kinetic energy, and $Pr_t$ is the turbulent Prandtl number. These relations create two additional unknowns, $\mu_u$ and $k_u$, which are determined through a modification of an underlying RANS turbulence model. In this work, the $k$-$\omega$ Shear Stress Transport (SST) model of \citet{Menter1994} is used. This model is adapted to the PANS formulation by reformulating the turbulence variables into their unresolved components, defined through the relations 
\begin{equation}
    f_k = \frac{k_u}{k} \quad \text{and} \quad f_\omega = \frac{\omega_u}{\omega},
\end{equation}
where the subscript $(\cdot)_u$ denotes the unresolved (i.e., modeled) component of the variable $(\cdot)$ and $f_{(\cdot)}$ denotes the unresolved-to-total ratio of $(\cdot)$. This formulation provides a mechanism for the PANS approach to seamlessly blend between DNS ($f_k = f_\omega = 0$) and URANS ($f_k = f_\omega = 1$). The modified transport equations for the unresolved turbulent kinetic energy $k_u$ and unresolved specific dissipation $\omega_u$ are given as
\begin{equation}
    \frac{\partial \rho k_u}{\partial t} + \frac{\partial}{\partial x_i}(\rho k_u u_i) 
    = P_k 
    - \beta^* \rho k_u \omega_u 
    + \frac{\partial}{\partial x_j}\bigg [\big(\mu + \mu_t \sigma_k \frac{f_\omega}{f_k}\big) \frac{\partial k_u}{\partial x_j} \bigg],
\end{equation}
\begin{multline}
    \frac{\partial \rho \omega_u}{\partial t} + \frac{\partial}{\partial x_i}(\rho \omega_u u_i) 
    = \frac{\alpha}{\mu_u} \rho P_k  
    - \rho \bigg (P' - \frac{P'}{f_\omega} + \frac{\beta \omega_u}{f_\omega} \bigg) 
    + \frac{\partial}{\partial x_j}\bigg [\big(\mu + \mu_t \sigma_\omega \frac{f_\omega}{f_k}\big) \frac{\partial \omega_u}{\partial x_j} \bigg] \\
    + 2 \rho \frac{\sigma_{\omega 2}}{\omega_u} \frac{f_\omega}{f_k} (1 - F_1) \frac{\partial k_u}{\partial x_j} \frac{\partial \omega_u}{\partial x_j},
\end{multline}
where $P_k = \tau_{ij} \frac{\partial u_i}{\partial x_j}$ and $P' = \rho \alpha \beta^* k_u/\mu_u$. The eddy viscosity is calculated as
\begin{equation}
    \mu_u = \frac{\rho \alpha_1 k_u}{\max (\alpha_1 \omega_u, \Omega F_2)},
\end{equation}
with the auxiliary functions defined as
\begin{align}
    F_1 &= \tanh \Biggr ( \min \bigg ( \max \bigg ( \frac{\sqrt{k_u}}{\beta^* \omega_u d}, \frac{500 \mu}{\rho \omega_u d^2} \bigg ),  \frac{4 \rho \sigma_{\omega 2} k_u}{CD_{k\omega} d^2}\bigg )^4 \Biggr),\\
    F_2 &= \tanh \biggr ( \max \bigg ( \frac{2 \sqrt{k_u}}{\beta^* \omega_u d}, \frac{500 \mu}{\rho \omega_u d^2} \bigg )^2 \biggr),\\
    CD_{k\omega} &= \max \bigg ( \frac{2 \rho \sigma_{\omega 2} }{\omega_u} \frac{\partial k_u}{\partial x_j} \frac{\partial \omega_u}{\partial x_j}, 10^{-10} \bigg ),\\
    \Omega &= \sqrt{2 W_{ij} W_{ij}}, \\
    W_{ij} &= \frac{1}{2} \bigg ( \frac{\partial u_i}{\partial x_j} - \frac{\partial u_j}{\partial x_i} \bigg ),
\end{align}
for some wall distance $d$. The values of the free parameters are tabulated in \cref{app:constants}, and description of the modifications applied to the PANS formulation to make it more amenable to high-order discretizations is presented in \cref{ssec:numerical}. 

The effectiveness of the PANS approach is ultimately reliant on the values of the resolution-control parameters $f_k$ and $f_\omega$. In this work, these parameters are kept constant in space and time as the presence of spatio-temporal variation in the parameters requires additional modifications in the governing equations to account for commutation errors. Furthermore, the two-parameter model ($f_k$, $f_\omega$) is reduced to a one-parameter model ($f_k$) through the assumption that the length scales associated with turbulent dissipation are entirely unresolved (i.e., $f_\epsilon =1$), an assumption proposed in \citet{Girimaji2013} and explored in \citet{Pereira2015}. Subsequently, the one-parameter model is closed through the relation $f_\omega = f_\epsilon/f_k = 1/f_k$.

\subsection{Spatial Discretization}

\begin{figure}[tbhp]
    \centering
  \hspace*{\fill}%
    \subfloat[$\mathbb P_1$]{
    \adjustbox{width=0.25\linewidth,valign=b}{     \begin{tikzpicture}[spy using outlines={rectangle, height=3cm,width=2.3cm, magnification=3, connect spies}]
		\begin{axis}[name=plot1,
		    axis line style={draw=none},
		    tick style={draw=none},
		    axis x line=left,
            axis y line=left,
            axis equal image,
            clip mode=individual,
    		xmin=-1,
    		xmax=1,
    		xticklabels={,,},
    		ymin=-1,
    		ymax=1,
    		yticklabels={,,},
    		style={font=\small},
    		scale = 1]

		    \draw[-] (axis cs:-1, -1) -- (axis cs:1, -1);
		    \draw[-] (axis cs:1, -1) -- (axis cs:-1, 1);
		    \draw[-] (axis cs:-1, 1) -- (axis cs:-1, -1);

            \draw[-,fill=red] (axis cs:-0.66666666666666666666666666666666666667, 0.33333333333333333333333333333333333333) circle[radius=0.04];
            \draw[-,fill=red] (axis cs:0.33333333333333333333333333333333333333, -0.66666666666666666666666666666666666667) circle[radius=0.04];
            \draw[-,fill=red] (axis cs:-0.66666666666666666666666666666666666667, -0.66666666666666666666666666666666666667) circle[radius=0.04];
            
            \draw [blue] plot [only marks, mark=square*, mark size=2.5] coordinates {
            (-1, -0.577350269189625764509148780502)
            (-1, 0.577350269189625764509148780502)
            (0.577350269189625764509148780502, -1)
            (-0.577350269189625764509148780502, -1)
            (-0.577350269189625764509148780502, 0.577350269189625764509148780502)
            (0.577350269189625764509148780502, -0.577350269189625764509148780502)};
            
		\end{axis}

	\end{tikzpicture}}}
    \subfloat[$\mathbb P_3$]{
    \adjustbox{width=0.25\linewidth,valign=b}{     \begin{tikzpicture}[spy using outlines={rectangle, height=3cm,width=2.3cm, magnification=3, connect spies}]
		\begin{axis}[name=plot1,
		    axis line style={draw=none},
		    tick style={draw=none},
		    axis x line=left,
            axis y line=left,
            axis equal image,
            clip mode=individual,
    		xmin=-1,
    		xmax=1,
    		xticklabels={,,},
    		ymin=-1,
    		ymax=1,
    		yticklabels={,,},
    		style={font=\small},
    		scale = 1]

		    \draw[-] (axis cs:-1, -1) -- (axis cs:1, -1);
		    \draw[-] (axis cs:1, -1) -- (axis cs:-1, 1);
		    \draw[-] (axis cs:-1, 1) -- (axis cs:-1, -1);

		    
            
            \draw[-,fill=red] (axis cs:-0.333333333333333,-0.333333333333333) circle[radius=0.04];
            \draw[-,fill=red] (axis cs:-0.888871894660414,0.777743789320828) circle[radius=0.04];
            \draw[-,fill=red] (axis cs:0.777743789320828,-0.888871894660414) circle[radius=0.04];
            \draw[-,fill=red] (axis cs:-0.888871894660414,-0.888871894660414) circle[radius=0.04];
            \draw[-,fill=red] (axis cs:0.268421495491446,-0.408932576528214) circle[radius=0.04];
            \draw[-,fill=red] (axis cs:-0.859488918963232,-0.408932576528214) circle[radius=0.04];
            \draw[-,fill=red] (axis cs:-0.408932576528214,-0.859488918963232) circle[radius=0.04];
            \draw[-,fill=red] (axis cs:0.268421495491446,-0.859488918963232) circle[radius=0.04];
            \draw[-,fill=red] (axis cs:-0.859488918963232,0.268421495491446) circle[radius=0.04];
            \draw[-,fill=red] (axis cs:-0.408932576528214,0.268421495491446) circle[radius=0.04];
            
            \draw [blue] plot [only marks, mark=square*, mark size=2.5] coordinates {
            (-1, -0.861136311594052575223946488893)
            (-1, -0.339981043584856264802665759103)
            (-1, 0.861136311594052575223946488893)
            (-1, 0.339981043584856264802665759103)
            (-0.861136311594052575223946488893, -1)
            (-0.339981043584856264802665759103, -1)
            (0.861136311594052575223946488893, -1)
            (0.339981043584856264802665759103, -1)
            (-0.861136311594052575223946488893, 0.861136311594052575223946488893)
            (-0.339981043584856264802665759103, 0.339981043584856264802665759103)
            (0.861136311594052575223946488893, -0.861136311594052575223946488893)
            (0.339981043584856264802665759103, -0.339981043584856264802665759103)};


		\end{axis}

	\end{tikzpicture}}}
  \hspace*{\fill}%
    \newline
    \caption{\label{fig:fr_elems} Diagram of the solution point (circles, $\mathbf{x}$) and interface flux point (squares, $\hat{\mathbf{x}}$) distributions for a $\mathbb P_1$ (left) and a $\mathbb P_3$ (right) triangle element using Williams-Shunn \citep{Williams2014} and Gauss-Legendre points, respectively. }
\end{figure}
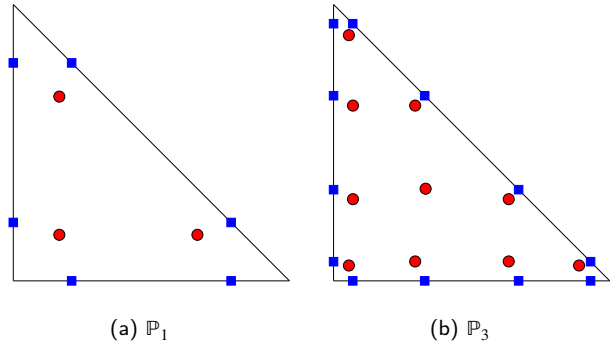

The PANS equations were discretized using the flux reconstruction approach of \citet{Huynh2007}, a generalization of the nodal discontinuous Galerkin method \citep{Hesthaven2008}. We present a brief description of the FR algorithm as it relates to first-order conservation laws, given in the form of 
\begin{equation}\label{eq:conlaw}
    \frac{\partial \mathbf{u}}{\partial t} + \boldsymbol{\nabla} \cdot \mathbf{F}(\mathbf{u}) = 0
\end{equation}
for a solution $\mathbf{u}(\mathbf{x},t)$ and flux $\mathbf{F}(\mathbf{u})$.
In this approach, the domain $(\mathbf{x},t)\in\Omega\times\mathbb{R}_+$ is partitioned into $N$ elements $\Omega_k$ such that $\Omega = \bigcup_N\Omega_k$ and $\Omega_i\cap\Omega_j=\emptyset$ for $i\neq j$. Within each element $\Omega_k$, shown in \cref{fig:fr_elems}, the approximate solution $\hat{\mathbf{u}}$ is represented by a discontinuous polynomial approximation of maximal order $p$, formed by a nodal interpolation through the set of $q$ unique nodes $\{\mathbf{x}_1,\dots,\mathbf{x}_{q}\}\in\Omega_k$ for $q \geq p+1$. This interpolation is represented by 
\begin{equation}
        \hat{\mathbf{u}} = \sum^{q}_{j=1} \mathbf{u}(\mathbf{x}_j)\phi_{j}(\mathbf{x}),
\end{equation}
where $\phi(\mathbf{x})$ is a set of nodal basis functions with the property that $\phi_{j}(\mathbf{x}_i) = \delta_{ij}$.
Similarly, $\hat{\mathbf{f}}^D$, a discontinuous approximation of the flux $\mathbf{F}(\hat{\mathbf{u}})$, is formed through the collocation projection of the flux at the solution nodes. 
\begin{equation}
        \hat{\mathbf{f}}^D = \sum^{q}_{j=1} \mathbf{F}(\mathbf{u}(x_j))\phi_{j}(\mathbf{x}).
\end{equation}

The approximate solution within each element is then evaluated at a set of $r$ interface points $\{\hat{\mathbf{x}}_1,\dots,\hat{\mathbf{x}}_{r}\}\in \partial \Omega_k$, such that at each interface point, these values, in conjunction with the analogous values from the neighboring elements, can be used to form a common interface flux $\mathbf{f}^I$. This is typically calculated by treating the interface value pairs as a Riemann problem using approaches such as those of \citet{Rusanov1962} and \citet{Roe1981}. A continuous flux function $\hat{\mathbf{f}}$ is then computed by adding correction terms, $\hat{\mathbf{f}}^C$, such that the flux evaluated at the interfaces equals the common flux. 
\begin{equation}
    \hat{\mathbf{f}} = \hat{\mathbf{f}}^D + \sum_{j = 1}^{r} \hat{\mathbf{f}}^C_j.
\end{equation}
These correction terms are defined as
\begin{equation}
    \hat{\mathbf{f}}^C_j = (\mathbf{f}^I_j - \hat{\mathbf{f}}^D_j) \mathbf{g}_j(\mathbf{x}),
\end{equation}
where $\mathbf{g}_j(\mathbf{x})$ denotes the $j$th correction function, $\mathbf{f}^I_j$ denotes the common interface flux at $\hat{\mathbf{x}}_j$, and $\hat{\mathbf{f}}^D_j$ denotes the discontinuous flux function evaluated at $\hat{\mathbf{x}}_j$. The correction functions have the property that 
\begin{equation*}
    \mathbf{n}_i \cdot \mathbf{g}_j(\mathbf{x}_i) = \delta_{ij} \quad \quad \mathrm{and} \quad \quad \sum_{j=1}^r \mathbf{g}_j(\mathbf{x}) \in \mathrm{RT}_p,
\end{equation*}
where $\mathbf{n}_i$ is the outward facing normal at $\mathbf{x}_i$ and $\mathrm{RT}_p$ is the Raviart-Thomas space of order $p$. The choice of these functions ultimately dictates the numerical properties of the scheme \citep{Huynh2007, Trojak2021}. The corrected flux $\hat{\mathbf{f}}$
can then be substituted into \cref{eq:conlaw} and integrated in time using a suitable temporal integration scheme.
For details on implementation and extensions to second-order systems, the reader is referred to the work of \citet{Witherden2016} and the references therein. Within this work, we use the notation $\mathbb{P}_p$ to denote a solution approximation with maximal order $p$.
 
\subsection{Numerical Implementation}\label{ssec:numerical}
Due to the inherently low numerical dissipation in high-order discretizations, several modifications were applied to the PANS equations to ensure robustness. In the context of the transport equations for the turbulence variables, ensuring that these variables remain positive requires adapting the formulation. For the $\omega_u$ transport equation, where the variable is \emph{strictly positive}, we take the approach of \citet{Ilinca1999} as described by \citet{Bassi2005}, in which the transport equation for $\tilde{\omega}_u = \log(\omega_u)$ is solved instead. 

\begin{multline}
    \frac{\partial \rho \tilde{\omega}_u}{\partial t} + \frac{\partial}{\partial x_i}(\rho \tilde{\omega}_u u_i) 
    = \frac{\alpha}{\mu_u} \rho P_k  
    - \rho \bigg (P' - \frac{P'}{f_\omega} + \frac{\beta e^{\tilde{\omega}_u}}{f_\omega} \bigg) 
    + \frac{\partial}{\partial x_j}\bigg [\big(\mu + \mu_t \sigma_\omega \frac{f_\omega}{f_k}\big) \frac{\partial \tilde{\omega}_u}{\partial x_j} \bigg] \\
    + 2 \rho \frac{\sigma_{\omega 2}}{e^{\tilde{\omega}_u}} \frac{f_\omega}{f_k} (1 - F_1) \frac{\partial k_u}{\partial x_j} \frac{\partial \tilde{\omega}_u}{\partial x_j}.
\end{multline}

As a result, the specific dissipation term only appears in exponential form in the transport and auxiliary equations, guaranteeing positivity. Furthermore, the distribution of the logarithm form of $\omega_u$ is smoother than that of $\omega_u$ itself \citep{Bassi2005}. However, for $k_u$, where the variable is only \emph{non-negative}, the logarithm form is not necessarily well-defined. In contrast to the approach of \citet{Bassi2005} where negative values of $k$ were allowed in the solution but limited to zero in the transport and auxiliary equations, we introduce a source term $S_k$ in the $k_u$ transport equation which activates if $k_u$ falls below some small constant $k_{min}$, such that $k_u$ is effectively limited to $k_{min}$. This source term was formed via a forward Euler approximation of $k_u$, given as
\begin{equation}
    S_k^{n} = \mathrm{max} \bigg [ 0,\ \frac{k_{min} - k_u^n}{\Delta t} + \boldsymbol \nabla \cdot \mathbf{F}_k^n \bigg ],
\end{equation}
where $\mathbf{F}_k$ is the $k_u$ component of the flux and the superscript $n$ denotes the time step. The value of $k_{min}$ was set to $10^{-8}$.

The PANS equations, along with these modifications, were implemented in the PyFR software package \citep{Witherden2014}. The equations were discretized with the FR approach using a Rusanov-type \citep{Rusanov1962} Riemann solver for the inviscid fluxes, the BR2 method of \citet{Bassi2000} for the viscous fluxes, and an explicit fourth-order Runge--Kutta scheme for temporal integration. In the numerical experiments, a comparison between a low-order and high-order FR approach was conducted. The low-order approach was performed using $\mathbb{P}_1$ solution polynomials, resulting in a second-order accurate scheme comparable to a finite volume formulation. For the high-order approach, $\mathbb{P}_3$ solution polynomials were used, resulting in a fourth-order accurate scheme. Due to the increase in resolution afforded through higher-order representations of the solution, the meshes used for the low-order and high-order approaches were coarsened/refined appropriately such that the total degrees of freedom remained the same (i.e., the $\mathbb{P}_1$ meshes used approximately 6-8 times as many elements as the $\mathbb{P}_3$ meshes). 

\section{Numerical Experiments}\label{sec:numerical}
The effects of a high-order discretization of the PANS equations were evaluated on two distinct numerical experiments. The first test case, the periodic hill problem of \citet{Frohlich2005}, consists of a recirculating, wall-bounded flow through a channel with periodic constrictions, a canonical benchmark for computing separated flows. The second test case, the flow around a circular cylinder at a Reynolds number of 3900, serves as an assessment of the methods for problems with significantly more complex flow physics, such as laminar separation, free-shear transition, and turbulent wakes. 
These assessments were performed at under-resolved to moderately-resolved levels of spatio-temporal resolution, where first-order statistics can be predicted reasonably well but second-order statistics and/or dominant flow physics may be poorly predicted. This range of resolution is where the application of hybrid turbulence models is most practical, as accurate predictions of the flow physics are achievable while still relaxing the resolution requirements for LES. As a result, the values of the free parameter $f_k$ were investigated over a range of $0.1$ to $0.3$. Furthermore, the PANS approach was compared to an under-resolved DNS (URDNS) approach for which the same mesh and numerical setup were used without the addition of the PANS model to give a comparable assessment of the effects of the model. The focus of these comparisons is towards metrics that are more difficult to resolve, such as second-order statistics, temporal properties of the flow physics, and the characteristics of the coherent structures in the flow. We reiterate that the goal of this investigation is not to evaluate the efficacy of the PANS method, but to show the effect of high-order discretizations for SRS using PANS and how that effect differs from approaches without any physical models.

\subsection{Periodic Hill}

\begin{figure}[tbhp]
    \centering
    \subfloat[Geometry]{
    \adjustbox{width=0.5\linewidth,valign=b}{     \begin{tikzpicture}[spy using outlines={rectangle, height=3cm,width=2.3cm, magnification=3, connect spies}]
		\begin{axis}[name=plot1,
		    axis line style={draw=none},
		    tick style={draw=none},
		    axis x line=left,
            axis y line=left,
            axis equal image,
            clip mode=individual,
    		xmin=0,
    		xmax=9,
    		xticklabels={,,},
    		ymin=0,
    		ymax=3,
    		yticklabels={,,},
    		style={font=\small},
    		scale = 1]
    		
			\addplot[name path=f, color=black, style={thick}]
			    table[x=x,y=y,col sep=comma,unbounded coords=jump]{./figs/data/periodichill/ph_geopoints.csv};
			    
            \path[name path=axis] (axis cs:0,0) -- (axis cs:9,0);
			    
            \addplot [
                thick,
                color=gray,
                fill=gray, 
                fill opacity=0.75
            ]
            fill between[
                of=f and axis,
                soft clip={domain=0:9},
            ];

		    \draw[-] (axis cs:0.0, 0.0) -- (axis cs:9.0, 0.0);
		    \draw[-] (axis cs:0.0, 0.0) -- (axis cs:0.0, 3.035);
		    \draw[-] (axis cs:9.0, 0.0) -- (axis cs:9.0, 3.035);
		    \draw[-] (axis cs:0.0, 3.035) -- (axis cs:9.0, 3.035);

		    \draw[<->] (axis cs:0.0, -0.2) -- (axis cs:9.0, -0.2);
		    \draw[<->] (axis cs:-0.2, 0) -- (axis cs:-0.2, 1);
		    \draw[<->] (axis cs:9.2, 0) -- (axis cs:9.2, 3.035);
		    
		    \node at (axis cs:4.5,-0.5) {$9h$};
		    \node at (axis cs:-0.5,0.5) {$h$};
		    \node at (axis cs:10.0,1.5) {$3.035h$};
			    
		\end{axis}

	\end{tikzpicture}}}
    \subfloat[Mesh]{
    \adjustbox{width=0.4\linewidth,valign=b}{\includegraphics[]{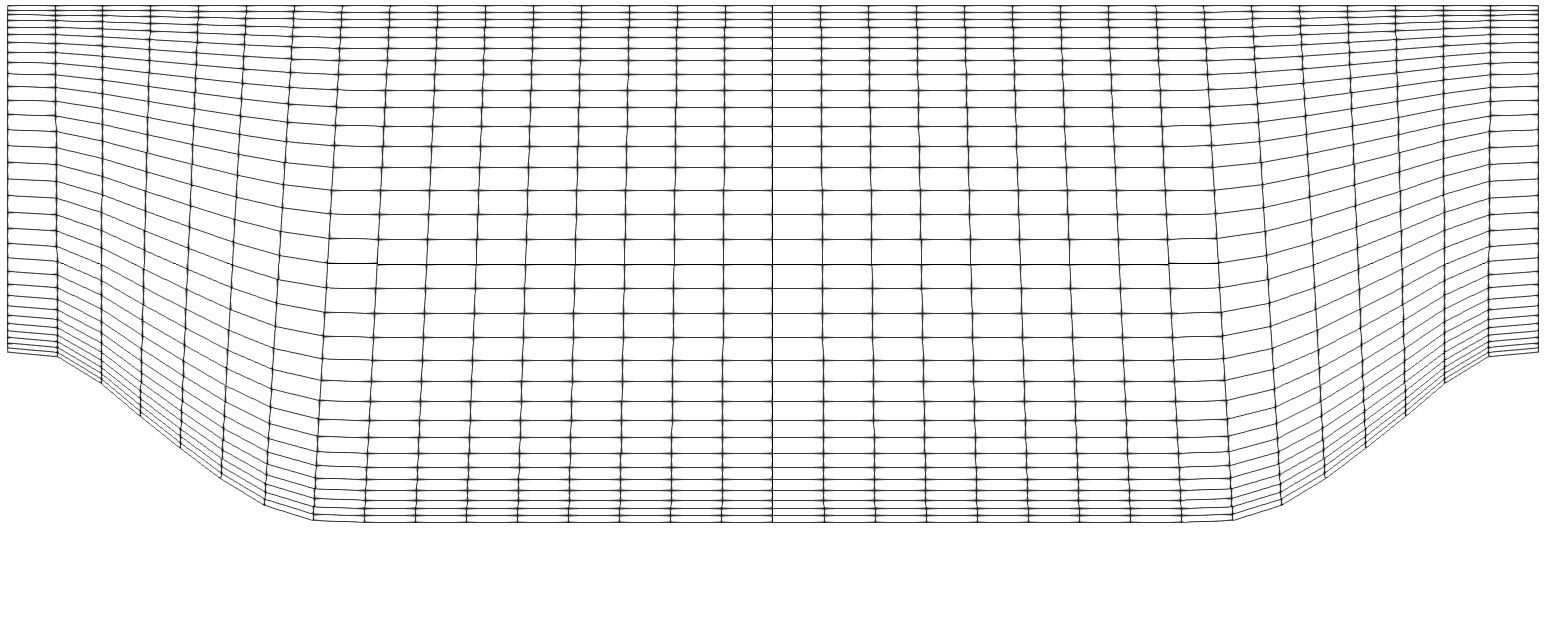}}}
    \newline
    \caption{\label{fig:ph_geo} Cross-section of the periodic hill geometry in the streamwise-spanwise plane (left) and mesh for the $\mathbb{P}_1$ approach (right). }
\end{figure}
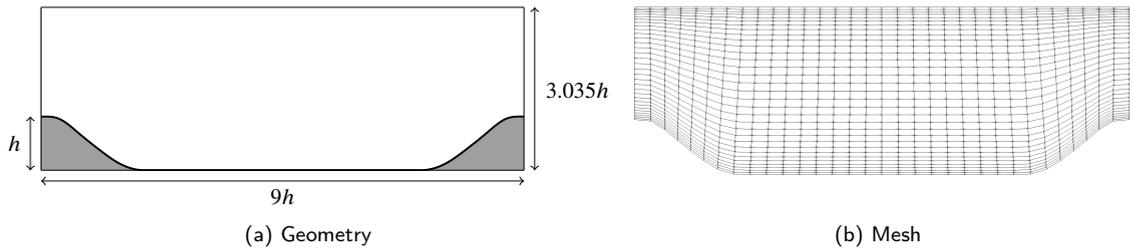

The periodic hill problem of \citet{Frohlich2005} presents a general assessment for predicting flow separation arising from curved surfaces and the subsequent flow reattachment. The problem consists of a channel flow with an infinite series of smooth constrictions (hills) of height $h$ separated by a crest-to-crest distance of $9h$. In numerical experiments, the infinite domain is approximated with a truncated region spanning one crest-to-crest distance with periodicity enforced between the inlet and outlet. A 2D cross-section of the model geometry in the streamwise-spanwise plane is shown in \cref{fig:ph_geo} for this truncated region. The 3D geometry is formed by extruding the cross-section along the spanwise direction over a length of $4.5h$.

Initially, a uniform flow with a Mach number of 0.1 is set throughout the domain, and the bulk Reynolds number, based on the bulk velocity $U_b$ and height $h$, is set to 10,595. To enforce a constant mass flow rate across the domain, a body force in the form of an additional streamwise pressure gradient was imposed using the approach of \citet{Benocci1990} as described by \citet{Wang2009}, given as
\begin{equation}
    \bigg ( \frac{dP}{dx} \bigg )^{n+1} = \bigg ( \frac{dP}{dx} \bigg )^{n} - \frac{1}{A_c \Delta t} (\Dot{m}^* - 2\Dot{m}^n + \Dot{m}^{n-1}),
\end{equation}
where the superscript denotes the time step iteration, $A_c$ is the inlet/outlet area, and $\Dot{m}$ is the mass flow rate at the inlet/outlet. The desired mass flow rate $\Dot{m}^*$ was set based on the bulk Reynolds number, and the initial pressure gradient was set to zero. 

Two structured, hexahedral meshes were generated for the periodic hill, one for the low-order ($\mathbb{P}_1$) approach and one for the high-order ($\mathbb{P}_3$) approach. The low-order mesh was generated by uniformly sub-dividing the high-order mesh along each direction. A description of the characteristics of these meshes is shown in \cref{tab:ph_grids}, where $N$ denotes the number of elements, $DOF$ denotes the degrees of freedom, and $x$, $y$, and $z$ denote the streamwise, normal, and spanwise directions, respectively. The $\Delta y^+$ value was estimated by normalizing the distance of the first solution point to the wall by the friction velocity of a flat plate at the bulk Reynolds number. The difference in $\Delta y^+$ values between the grids results from the nonuniformity of the solution point distribution at various orders. To highlight the differences in the predicted flow properties between the low-order and high-order approaches, significantly coarser grids were used in the present work, using 97\% fewer degrees of freedom than the LES of \citet{Frohlich2005} and 85\% fewer degrees of freedom than the low-order PANS approach of \citet{Razi2017}. As such, the results are presented for $f_k$ values of 0.2 and 0.3, as through \emph{a posteriori} analysis, it was shown that this resolution level does not support lower $f_k$ values.

\begin{figure}[tbhp]
    \centering
    \begin{tabular}{r c c c c c}\toprule
        Grid & Method & Order & $N_x \times N_y \times N_z$ & $DOF_x \times DOF_y \times DOF_z$ & $\Delta y^+$  \\ \midrule
        
        A1 &
        PANS/URDNS &
        $\mathbb{P}_1$ &
        $32 \times 32 \times 16$ &
        $64 \times 64 \times 32$ &
        1.8 \\
        
        A2 &  
        PANS/URDNS &
        $\mathbb{P}_3$ &
        $16 \times 16 \times 8$ &
        $64 \times 64 \times 32$ &
        1.0\\ 
        \bottomrule
    \end{tabular}
    
    \captionof{table}{\label{tab:ph_grids}Description of mesh characteristics for the periodic hill case. }
\end{figure}

Periodic boundary conditions were imposed between the inlet and outlet as well as along the spanwise direction. At both the top and bottom walls, no-slip, adiabatic boundary conditions were applied. Following the recommendation of \citet{Menter1994}, the boundary conditions at the wall for the turbulence variables were set as
\begin{equation}\label{eq:pansbcs}
    k_u = 0, \quad \quad \omega_u = f_\omega \frac{60 \nu}{\beta_1 (\Delta d_1)^2},
\end{equation}
where $\beta_1$ is a constant defined in \cref{app:constants} and $\Delta d_1$ is the wall distance. To maintain consistency in the boundary conditions between low-order and high-order approaches in light of the nonuniformity of the solution point distribution, the wall distance was approximated as $\Delta d_1 = \Delta d_e/(p+1)$, where $\Delta d_e$ is the height of the element and $p$ is the order of the solution basis. As such, the values of the turbulence variables at the wall were identical between both approaches. The simulation was run for a period corresponding to 20 flow-through times of the domain. After 10 flow-through times, the flow was assumed to be fully-developed. Statistical quantities were then gathered over the final 10 flow-through times.

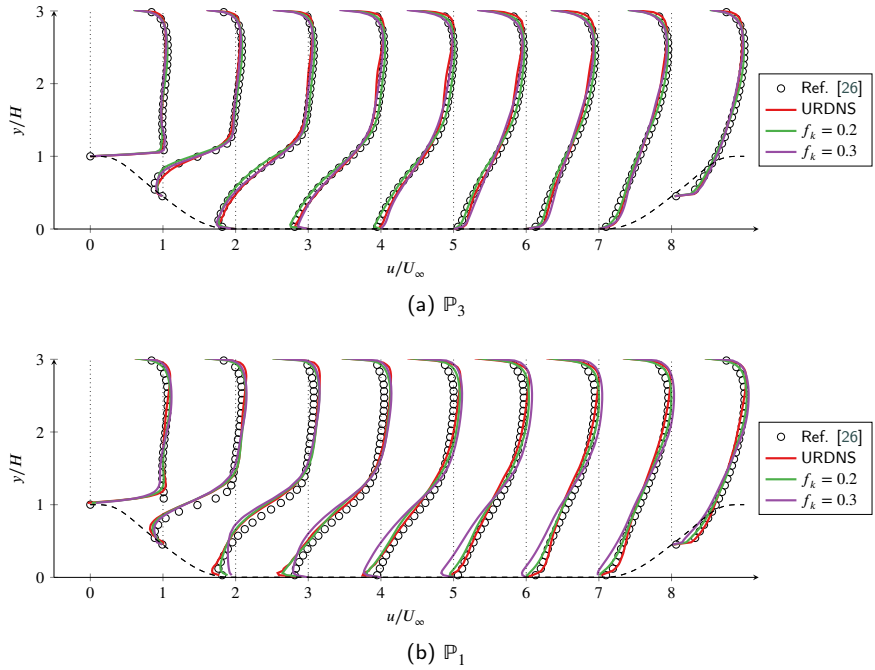
\begin{figure}[tbhp]
        \centering
        \subfloat[$\mathbb{P}_3$]{\label{fig:ph_u_h}
        \adjustbox{width=0.7\linewidth,valign=b}{     \begin{tikzpicture}[spy using outlines={rectangle, height=3cm,width=2.3cm, magnification=3, connect spies}]
		\begin{axis}[name=plot1,
		    axis line style={latex-latex},
		    axis x line=left,
            axis y line=left,
            axis equal image,
            clip mode=individual,
		    xlabel={$u/U_\infty$},
		    xtick={0, 1, 2, 3, 4, 5, 6, 7, 8},
    		xmin=-0.5,
    		xmax=9.2,
    		x tick label style={
        		/pgf/number format/.cd,
            	fixed,
            	precision=1,
        	    /tikz/.cd},
    		ylabel={$y/H$},
    		ytick={0, 1, 2, 3},
    		ymin=0,
    		ymax=3,
    		y tick label style={
        		/pgf/number format/.cd,
            	fixed,
            	precision=1,
        	    /tikz/.cd},
    		legend style={at={(1.0,0.5)},anchor=west},
    		legend cell align={left},
    		style={font=\small},
    		scale = 2]

            \def\x{0};
    			\addplot[color=black, style={ultra thin}, only marks, mark=o, mark options={scale=1.0}, mark repeat = 3, mark phase = 0]
    				table[x expr={\thisrow{u}+ \x},y = y/h,col sep=comma,unbounded coords=jump]{./figs/data/periodichill/ref_x\x.csv};
    		    \addlegendentry{Ref. \citep{Frohlich2005}}
    		    
    			\addplot[color={Set1-A}, style={very thick}]
    				table[x expr={\thisrow{u}+ \x},y = y/h,col sep=comma,unbounded coords=jump]{./figs/data/periodichill/LES_medp3_x\x.csv};
        		\addlegendentry{URDNS}

    		
    			\addplot[color={Set1-C}, style={very thick}]
    				table[x expr={\thisrow{u}+ \x},y = y/h,col sep=comma,unbounded coords=jump]{./figs/data/periodichill/fk02_medp3_x\x.csv};
        		\addlegendentry{$f_k = 0.2$}
    		
    			\addplot[color={Set1-D}, style={ very thick}]
    				table[x expr={\thisrow{u}+ \x},y = y/h,col sep=comma,unbounded coords=jump]{./figs/data/periodichill/fk03_medp3_x\x.csv};
        		\addlegendentry{$f_k = 0.3$}
    		
            \foreach \x in {1,...,8}{
                \addplot[color=black, style={ultra thin}, only marks, mark=o, mark options={scale=1.0}, mark repeat = 3, mark phase = 0]
    				table[x expr={\thisrow{u}+ \x},y = y/h,col sep=comma,unbounded coords=jump]{./figs/data/periodichill/ref_x\x.csv};
    				
    			\addplot[color={Set1-A}, style={very thick}]
    				table[x expr={\thisrow{u}+ \x},y = y/h,col sep=comma,unbounded coords=jump]{./figs/data/periodichill/LES_medp3_x\x.csv};

    		
    			\addplot[color={Set1-C}, style={very thick}]
    				table[x expr={\thisrow{u}+ \x},y = y/h,col sep=comma,unbounded coords=jump]{./figs/data/periodichill/fk02_medp3_x\x.csv};
    		
    			\addplot[color={Set1-D}, style={very thick}]
    				table[x expr={\thisrow{u}+ \x},y = y/h,col sep=comma,unbounded coords=jump]{./figs/data/periodichill/fk03_medp3_x\x.csv};
            }
    		
			\addplot[color=black, style={thin}, dashed]
			    table[x=x,y=y,col sep=comma,unbounded coords=jump]{./figs/data/periodichill/ph_geopoints.csv};
			\addplot[color=black, style={thin}, dashed]
			    table[x=x,y=y,col sep=comma,unbounded coords=jump]{./figs/data/periodichill/ph_geopoints.csv};
			    
    		\draw [dotted] (0,1) -- (0,3);
    		\draw [dotted] (1,.45) -- (1,3);
    		\draw [dotted] (2,0) -- (2,3);
    		\draw [dotted] (3,0) -- (3,3);
    		\draw [dotted] (4,0) -- (4,3);
    		\draw [dotted] (5,0) -- (5,3);
    		\draw [dotted] (6,0) -- (6,3);
    		\draw [dotted] (7,0) -- (7,3);
    		\draw [dotted] (8,.45) -- (8,3);
		\end{axis}

	\end{tikzpicture}}}
        \newline
        \subfloat[$\mathbb{P}_1$]{\label{fig:ph_u_l}
        \adjustbox{width=0.7\linewidth,valign=b}{     \begin{tikzpicture}[spy using outlines={rectangle, height=3cm,width=2.3cm, magnification=3, connect spies}]
		\begin{axis}[name=plot1,
		    axis line style={latex-latex},
		    axis x line=left,
            axis y line=left,
            axis equal image,
            clip mode=individual,
		    xlabel={$u/U_\infty$},
		    xtick={0, 1, 2, 3, 4, 5, 6, 7, 8},
    		xmin=-0.5,
    		xmax=9.2,
    		x tick label style={
        		/pgf/number format/.cd,
            	fixed,
            	precision=1,
        	    /tikz/.cd},
    		ylabel={$y/H$},
    		ytick={0, 1, 2, 3},
    		ymin=0,
    		ymax=3,
    		y tick label style={
        		/pgf/number format/.cd,
            	fixed,
            	precision=1,
        	    /tikz/.cd},
    		legend style={at={(1.0,0.5)},anchor=west},
    		legend cell align={left},
    		style={font=\small},
    		scale = 2]

            \def\x{0};
    			\addplot[color=black, style={ultra thin}, only marks, mark=o, mark options={scale=1.0}, mark repeat = 3, mark phase = 0]
    				table[x expr={\thisrow{u}+ \x},y = y/h,col sep=comma,unbounded coords=jump]{./figs/data/periodichill/ref_x\x.csv};
    		    \addlegendentry{Ref. \citep{Frohlich2005}}
    		    
    			\addplot[color={Set1-A}, style={very thick}]
    				table[x expr={\thisrow{u}+ \x},y = y/h,col sep=comma,unbounded coords=jump]{./figs/data/periodichill/LES_finep1_x\x.csv};
        		\addlegendentry{URDNS}

    		
    			\addplot[color={Set1-C}, style={very thick}]
    				table[x expr={\thisrow{u}+ \x},y = y/h,col sep=comma,unbounded coords=jump]{./figs/data/periodichill/fk02_finep1_x\x.csv};
        		\addlegendentry{$f_k = 0.2$}
    		
    			\addplot[color={Set1-D}, style={ very thick}]
    				table[x expr={\thisrow{u}+ \x},y = y/h,col sep=comma,unbounded coords=jump]{./figs/data/periodichill/fk03_finep1_x\x.csv};
        		\addlegendentry{$f_k = 0.3$}
    		
            \foreach \x in {1,...,8}{
                \addplot[color=black, style={ultra thin}, only marks, mark=o, mark options={scale=1.0}, mark repeat = 3, mark phase = 0]
    				table[x expr={\thisrow{u}+ \x},y = y/h,col sep=comma,unbounded coords=jump]{./figs/data/periodichill/ref_x\x.csv};
    				
    			\addplot[color={Set1-A}, style={very thick}]
    				table[x expr={\thisrow{u}+ \x},y = y/h,col sep=comma,unbounded coords=jump]{./figs/data/periodichill/LES_finep1_x\x.csv};

    		
    			\addplot[color={Set1-C}, style={very thick}]
    				table[x expr={\thisrow{u}+ \x},y = y/h,col sep=comma,unbounded coords=jump]{./figs/data/periodichill/fk02_finep1_x\x.csv};
    		
    			\addplot[color={Set1-D}, style={very thick}]
    				table[x expr={\thisrow{u}+ \x},y = y/h,col sep=comma,unbounded coords=jump]{./figs/data/periodichill/fk03_finep1_x\x.csv};
            }
    		
			\addplot[color=black, style={thin}, dashed]
			    table[x=x,y=y,col sep=comma,unbounded coords=jump]{./figs/data/periodichill/ph_geopoints.csv};
			\addplot[color=black, style={thin}, dashed]
			    table[x=x,y=y,col sep=comma,unbounded coords=jump]{./figs/data/periodichill/ph_geopoints.csv};
			    
    		\draw [dotted] (0,1) -- (0,3);
    		\draw [dotted] (1,.45) -- (1,3);
    		\draw [dotted] (2,0) -- (2,3);
    		\draw [dotted] (3,0) -- (3,3);
    		\draw [dotted] (4,0) -- (4,3);
    		\draw [dotted] (5,0) -- (5,3);
    		\draw [dotted] (6,0) -- (6,3);
    		\draw [dotted] (7,0) -- (7,3);
    		\draw [dotted] (8,.45) -- (8,3);
		\end{axis}

	\end{tikzpicture}}}
        \newline
        \caption{\label{fig:ph_u}Time and span-averaged streamwise velocity profiles using a $\mathbb{P}_3$ (top) and $\mathbb{P}_1$ FR scheme (bottom). Profiles are shifted by 0, +1, +2, ..., +8, respectively, along the abscissa. }
    \end{figure}

After averaging the flow across the time-averaging horizon and along the spanwise direction, the first-order and second-order statistics of the periodic hill flow were analyzed. These quantities were compared to the LES results of \citet{Frohlich2005}. 
A comparison of the profiles of the average streamwise velocity as predicted by the $\mathbb{P}_1$ and $\mathbb{P}_3$ methods is shown in \cref{fig:ph_u}. For both methods, the separation of the flow aft of the hill was evident, but discrepancies in the profiles in the separation region were observed. For the low-order method, both the PANS and URDNS approaches overpredicted the reversal of the flow in the separation region, which resulted in notable deviations from the reference data in the region $0 \leq x \leq 5$. The introduction of the PANS model did not significantly improve the low-order results. However, for the high-order scheme, both the URDNS and PANS approaches performed significantly better, and the $f_k =0.2$ results were in excellent agreement with the reference data. Less deviation in the mean velocity profiles between the two values of $f_k$ were observed with the high-order scheme, indicating that the high-order approach is less sensitive to the application of the PANS model than the low-order approach. 

    \begin{figure}[tbhp]
        \centering
        \subfloat[$\mathbb{P}_3$]{\label{fig:ph_uu_h}
        \adjustbox{width=0.7\linewidth,valign=b}{     \begin{tikzpicture}[spy using outlines={rectangle, height=3cm,width=2.3cm, magnification=3, connect spies}]
		\begin{axis}[name=plot1,
		    axis line style={latex-latex},
		    axis x line=left,
            axis y line=left,
            axis equal image,
            clip mode=individual,
		    xlabel={$u'u'/U_\infty^2$},
		    xtick={0, 1, 2, 3, 4, 5, 6, 7, 8},
    		xmin=-0.5,
    		xmax=9.2,
    		x tick label style={
        		/pgf/number format/.cd,
            	fixed,
            	precision=1,
        	    /tikz/.cd},
    		ylabel={$y/H$},
    		ytick={0, 1, 2, 3},
    		ymin=0,
    		ymax=3,
    		y tick label style={
        		/pgf/number format/.cd,
            	fixed,
            	precision=1,
        	    /tikz/.cd},
    		legend style={at={(1.0,0.5)},anchor=west},
    		legend cell align={left},
    		style={font=\small},
    		scale = 2]

            \def\x{0};
    			\addplot[color=black, style={ultra thin}, only marks, mark=o, mark options={scale=1.0}, mark repeat = 3, mark phase = 0]
    				table[x expr={5*\thisrow{uu}+ \x},y = y/h,col sep=comma,unbounded coords=jump]{./figs/data/periodichill/ref_x\x.csv};
    		    \addlegendentry{Ref. \citep{Frohlich2005}}
    		    
    			\addplot[color={Set1-A}, style={very thick}]
    				table[x expr={5*\thisrow{uu}+ \x},y = y/h,col sep=comma,unbounded coords=jump]{./figs/data/periodichill/LES_medp3_x\x.csv};
        		\addlegendentry{URDNS}

    		
    			\addplot[color={Set1-C}, style={very thick}]
    				table[x expr={5*\thisrow{uu}+ \x},y = y/h,col sep=comma,unbounded coords=jump]{./figs/data/periodichill/fk02_medp3_x\x.csv};
        		\addlegendentry{$f_k = 0.2$}
    		
    			\addplot[color={Set1-D}, style={ very thick}]
    				table[x expr={5*\thisrow{uu}+ \x},y = y/h,col sep=comma,unbounded coords=jump]{./figs/data/periodichill/fk03_medp3_x\x.csv};
        		\addlegendentry{$f_k = 0.3$}
    		
            \foreach \x in {1,...,8}{
                \addplot[color=black, style={ultra thin}, only marks, mark=o, mark options={scale=1.0}, mark repeat = 3, mark phase = 0]
    				table[x expr={5*\thisrow{uu}+ \x},y = y/h,col sep=comma,unbounded coords=jump]{./figs/data/periodichill/ref_x\x.csv};
    				
    			\addplot[color={Set1-A}, style={very thick}]
    				table[x expr={5*\thisrow{uu}+ \x},y = y/h,col sep=comma,unbounded coords=jump]{./figs/data/periodichill/LES_medp3_x\x.csv};

    		
    			\addplot[color={Set1-C}, style={very thick}]
    				table[x expr={5*\thisrow{uu}+ \x},y = y/h,col sep=comma,unbounded coords=jump]{./figs/data/periodichill/fk02_medp3_x\x.csv};
    		
    			\addplot[color={Set1-D}, style={very thick}]
    				table[x expr={5*\thisrow{uu}+ \x},y = y/h,col sep=comma,unbounded coords=jump]{./figs/data/periodichill/fk03_medp3_x\x.csv};
            }
    		
			\addplot[color=black, style={thin}, dashed]
			    table[x=x,y=y,col sep=comma,unbounded coords=jump]{./figs/data/periodichill/ph_geopoints.csv};
			\addplot[color=black, style={thin}, dashed]
			    table[x=x,y=y,col sep=comma,unbounded coords=jump]{./figs/data/periodichill/ph_geopoints.csv};
			    
    		\draw [dotted] (0,1) -- (0,3);
    		\draw [dotted] (1,.45) -- (1,3);
    		\draw [dotted] (2,0) -- (2,3);
    		\draw [dotted] (3,0) -- (3,3);
    		\draw [dotted] (4,0) -- (4,3);
    		\draw [dotted] (5,0) -- (5,3);
    		\draw [dotted] (6,0) -- (6,3);
    		\draw [dotted] (7,0) -- (7,3);
    		\draw [dotted] (8,.45) -- (8,3);
		\end{axis}

	\end{tikzpicture}}}
        \newline
        \subfloat[$\mathbb{P}_1$]{\label{fig:ph_uu_l}
        \adjustbox{width=0.7\linewidth,valign=b}{     \begin{tikzpicture}[spy using outlines={rectangle, height=3cm,width=2.3cm, magnification=3, connect spies}]
		\begin{axis}[name=plot1,
		    axis line style={latex-latex},
		    axis x line=left,
            axis y line=left,
            axis equal image,
            clip mode=individual,
		    xlabel={$u'u'/U_\infty^2$},
		    xtick={0, 1, 2, 3, 4, 5, 6, 7, 8},
    		xmin=-0.5,
    		xmax=9.2,
    		x tick label style={
        		/pgf/number format/.cd,
            	fixed,
            	precision=1,
        	    /tikz/.cd},
    		ylabel={$y/H$},
    		ytick={0, 1, 2, 3},
    		ymin=0,
    		ymax=3,
    		y tick label style={
        		/pgf/number format/.cd,
            	fixed,
            	precision=1,
        	    /tikz/.cd},
    		legend style={at={(1.0,0.5)},anchor=west},
    		legend cell align={left},
    		style={font=\small},
    		scale = 2]

            \def\x{0};
    			\addplot[color=black, style={ultra thin}, only marks, mark=o, mark options={scale=1.0}, mark repeat = 3, mark phase = 0]
    				table[x expr={5*\thisrow{uu}+ \x},y = y/h,col sep=comma,unbounded coords=jump]{./figs/data/periodichill/ref_x\x.csv};
    		    \addlegendentry{Ref. \citep{Frohlich2005}}
    		    
    			\addplot[color={Set1-A}, style={very thick}]
    				table[x expr={5*\thisrow{uu}+ \x},y = y/h,col sep=comma,unbounded coords=jump]{./figs/data/periodichill/LES_finep1_x\x.csv};
        		\addlegendentry{URDNS}

    		
    			\addplot[color={Set1-C}, style={very thick}]
    				table[x expr={5*\thisrow{uu}+ \x},y = y/h,col sep=comma,unbounded coords=jump]{./figs/data/periodichill/fk02_finep1_x\x.csv};
        		\addlegendentry{$f_k = 0.2$}
    		
    			\addplot[color={Set1-D}, style={ very thick}]
    				table[x expr={5*\thisrow{uu}+ \x},y = y/h,col sep=comma,unbounded coords=jump]{./figs/data/periodichill/fk03_finep1_x\x.csv};
        		\addlegendentry{$f_k = 0.3$}
    		
            \foreach \x in {1,...,8}{
                \addplot[color=black, style={ultra thin}, only marks, mark=o, mark options={scale=1.0}, mark repeat = 3, mark phase = 0]
    				table[x expr={5*\thisrow{uu}+ \x},y = y/h,col sep=comma,unbounded coords=jump]{./figs/data/periodichill/ref_x\x.csv};
    				
    			\addplot[color={Set1-A}, style={very thick}]
    				table[x expr={5*\thisrow{uu}+ \x},y = y/h,col sep=comma,unbounded coords=jump]{./figs/data/periodichill/LES_finep1_x\x.csv};

    		
    			\addplot[color={Set1-C}, style={very thick}]
    				table[x expr={5*\thisrow{uu}+ \x},y = y/h,col sep=comma,unbounded coords=jump]{./figs/data/periodichill/fk02_finep1_x\x.csv};
    		
    			\addplot[color={Set1-D}, style={very thick}]
    				table[x expr={5*\thisrow{uu}+ \x},y = y/h,col sep=comma,unbounded coords=jump]{./figs/data/periodichill/fk03_finep1_x\x.csv};
            }
    		
			\addplot[color=black, style={thin}, dashed]
			    table[x=x,y=y,col sep=comma,unbounded coords=jump]{./figs/data/periodichill/ph_geopoints.csv};
			\addplot[color=black, style={thin}, dashed]
			    table[x=x,y=y,col sep=comma,unbounded coords=jump]{./figs/data/periodichill/ph_geopoints.csv};
			    
    		\draw [dotted] (0,1) -- (0,3);
    		\draw [dotted] (1,.45) -- (1,3);
    		\draw [dotted] (2,0) -- (2,3);
    		\draw [dotted] (3,0) -- (3,3);
    		\draw [dotted] (4,0) -- (4,3);
    		\draw [dotted] (5,0) -- (5,3);
    		\draw [dotted] (6,0) -- (6,3);
    		\draw [dotted] (7,0) -- (7,3);
    		\draw [dotted] (8,.45) -- (8,3);
		\end{axis}

	\end{tikzpicture}}}
        \newline
        \caption{\label{fig:ph_uu}Time and span-averaged streamwise velocity variance profiles using a $\mathbb{P}_3$ (top) and $\mathbb{P}_1$ FR scheme (bottom). Profiles are scaled by a factor of 5 and shifted by 0, +1, +2, ..., +8, respectively, along the abscissa. }
    \end{figure}
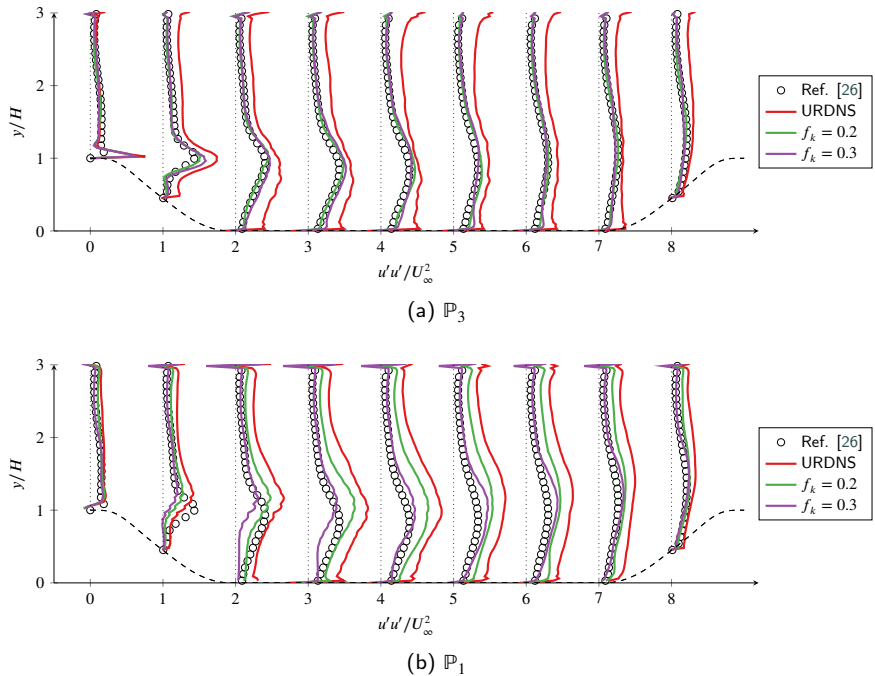
For the second-order statistics, the differences in the predictions by the various approaches were much more evident. In \cref{fig:ph_uu}, the profiles of the streamwise velocity variance are shown. For both the low-order and high-order methods, the URDNS approach significantly overpredicted the variance across most of the domain, with the largest overprediction seen in the separation region. The change from the low-order to high-order method for the URDNS approach did not appreciably improve the results. However, unlike URDNS, notable improvement was seen with the PANS approach when switching from the low-order method to the high-order method. Excellent agreement was observed between the high-order PANS approach and the reference data for both values of $f_k$, whereas the accuracy of the low-order PANS approach was comparable to the URDNS approaches. As with the first-order statistics, significantly less deviation in the results was observed between the two $f_k$ values when using the high-order scheme.
    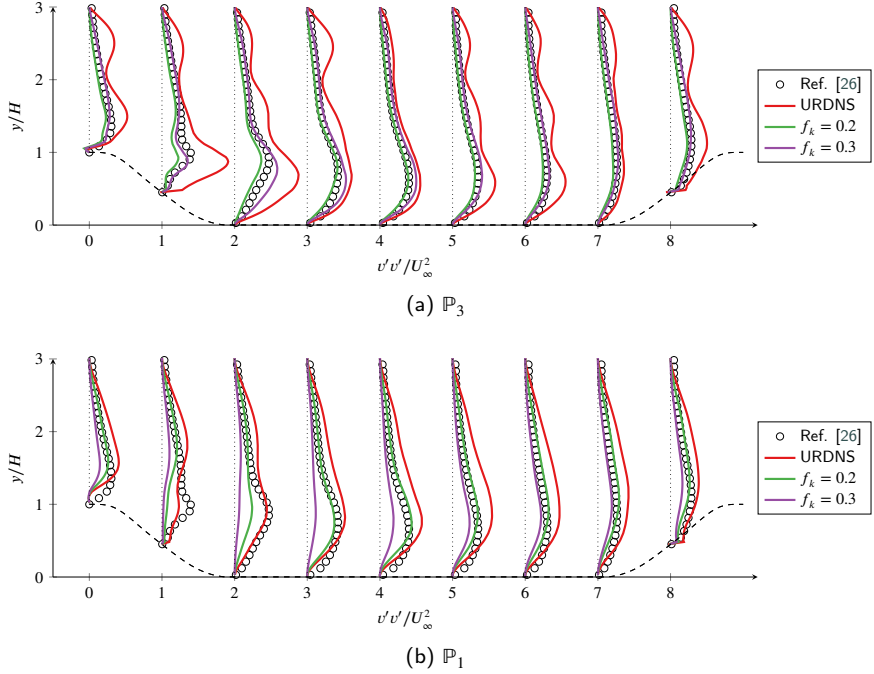
\begin{figure}[tbhp]
        \centering
        \subfloat[$\mathbb{P}_3$]{\label{fig:ph_vv_h}
        \adjustbox{width=0.7\linewidth,valign=b}{     \begin{tikzpicture}[spy using outlines={rectangle, height=3cm,width=2.3cm, magnification=3, connect spies}]
		\begin{axis}[name=plot1,
		    axis line style={latex-latex},
		    axis x line=left,
            axis y line=left,
            axis equal image,
            clip mode=individual,
		    xlabel={$v'v'/U_\infty^2$},
		    xtick={0, 1, 2, 3, 4, 5, 6, 7, 8},
    		xmin=-0.5,
    		xmax=9.2,
    		x tick label style={
        		/pgf/number format/.cd,
            	fixed,
            	precision=1,
        	    /tikz/.cd},
    		ylabel={$y/H$},
    		ytick={0, 1, 2, 3},
    		ymin=0,
    		ymax=3,
    		y tick label style={
        		/pgf/number format/.cd,
            	fixed,
            	precision=1,
        	    /tikz/.cd},
    		legend style={at={(1.0,0.5)},anchor=west},
    		legend cell align={left},
    		style={font=\small},
    		scale = 2]

            \def\x{0};
    			\addplot[color=black, style={ultra thin}, only marks, mark=o, mark options={scale=1.0}, mark repeat = 3, mark phase = 0]
    				table[x expr={10*\thisrow{vv}+ \x},y = y/h,col sep=comma,unbounded coords=jump]{./figs/data/periodichill/ref_x\x.csv};
    		    \addlegendentry{Ref. \citep{Frohlich2005}}
    		    
    			\addplot[color={Set1-A}, style={very thick}]
    				table[x expr={10*\thisrow{vv}+ \x},y = y/h,col sep=comma,unbounded coords=jump]{./figs/data/periodichill/LES_medp3_x\x.csv};
        		\addlegendentry{URDNS}

    		
    			\addplot[color={Set1-C}, style={very thick}]
    				table[x expr={10*\thisrow{vv}+ \x},y = y/h,col sep=comma,unbounded coords=jump]{./figs/data/periodichill/fk02_medp3_x\x.csv};
        		\addlegendentry{$f_k = 0.2$}
    		
    			\addplot[color={Set1-D}, style={ very thick}]
    				table[x expr={10*\thisrow{vv}+ \x},y = y/h,col sep=comma,unbounded coords=jump]{./figs/data/periodichill/fk03_medp3_x\x.csv};
        		\addlegendentry{$f_k = 0.3$}
    		
            \foreach \x in {1,...,8}{
                \addplot[color=black, style={ultra thin}, only marks, mark=o, mark options={scale=1.0}, mark repeat = 3, mark phase = 0]
    				table[x expr={10*\thisrow{vv}+ \x},y = y/h,col sep=comma,unbounded coords=jump]{./figs/data/periodichill/ref_x\x.csv};
    				
    			\addplot[color={Set1-A}, style={very thick}]
    				table[x expr={10*\thisrow{vv}+ \x},y = y/h,col sep=comma,unbounded coords=jump]{./figs/data/periodichill/LES_medp3_x\x.csv};

    		
    			\addplot[color={Set1-C}, style={very thick}]
    				table[x expr={10*\thisrow{vv}+ \x},y = y/h,col sep=comma,unbounded coords=jump]{./figs/data/periodichill/fk02_medp3_x\x.csv};
    		
    			\addplot[color={Set1-D}, style={very thick}]
    				table[x expr={10*\thisrow{vv}+ \x},y = y/h,col sep=comma,unbounded coords=jump]{./figs/data/periodichill/fk03_medp3_x\x.csv};
            }
    		
			\addplot[color=black, style={thin}, dashed]
			    table[x=x,y=y,col sep=comma,unbounded coords=jump]{./figs/data/periodichill/ph_geopoints.csv};
			\addplot[color=black, style={thin}, dashed]
			    table[x=x,y=y,col sep=comma,unbounded coords=jump]{./figs/data/periodichill/ph_geopoints.csv};
			    
    		\draw [dotted] (0,1) -- (0,3);
    		\draw [dotted] (1,.45) -- (1,3);
    		\draw [dotted] (2,0) -- (2,3);
    		\draw [dotted] (3,0) -- (3,3);
    		\draw [dotted] (4,0) -- (4,3);
    		\draw [dotted] (5,0) -- (5,3);
    		\draw [dotted] (6,0) -- (6,3);
    		\draw [dotted] (7,0) -- (7,3);
    		\draw [dotted] (8,.45) -- (8,3);
		\end{axis}

	\end{tikzpicture}}}
        \newline
        \subfloat[$\mathbb{P}_1$]{\label{fig:ph_vv_l}
        \adjustbox{width=0.7\linewidth,valign=b}{     \begin{tikzpicture}[spy using outlines={rectangle, height=3cm,width=2.3cm, magnification=3, connect spies}]
		\begin{axis}[name=plot1,
		    axis line style={latex-latex},
		    axis x line=left,
            axis y line=left,
            axis equal image,
            clip mode=individual,
		    xlabel={$v'v'/U_\infty^2$},
		    xtick={0, 1, 2, 3, 4, 5, 6, 7, 8},
    		xmin=-0.5,
    		xmax=9.2,
    		x tick label style={
        		/pgf/number format/.cd,
            	fixed,
            	precision=1,
        	    /tikz/.cd},
    		ylabel={$y/H$},
    		ytick={0, 1, 2, 3},
    		ymin=0,
    		ymax=3,
    		y tick label style={
        		/pgf/number format/.cd,
            	fixed,
            	precision=1,
        	    /tikz/.cd},
    		legend style={at={(1.0,0.5)},anchor=west},
    		legend cell align={left},
    		style={font=\small},
    		scale = 2]

            \def\x{0};
    			\addplot[color=black, style={ultra thin}, only marks, mark=o, mark options={scale=1.0}, mark repeat = 3, mark phase = 0]
    				table[x expr={10*\thisrow{vv}+ \x},y = y/h,col sep=comma,unbounded coords=jump]{./figs/data/periodichill/ref_x\x.csv};
    		    \addlegendentry{Ref. \citep{Frohlich2005}}
    		    
    			\addplot[color={Set1-A}, style={very thick}]
    				table[x expr={10*\thisrow{vv}+ \x},y = y/h,col sep=comma,unbounded coords=jump]{./figs/data/periodichill/LES_finep1_x\x.csv};
        		\addlegendentry{URDNS}

    		
    			\addplot[color={Set1-C}, style={very thick}]
    				table[x expr={10*\thisrow{vv}+ \x},y = y/h,col sep=comma,unbounded coords=jump]{./figs/data/periodichill/fk02_finep1_x\x.csv};
        		\addlegendentry{$f_k = 0.2$}
    		
    			\addplot[color={Set1-D}, style={ very thick}]
    				table[x expr={10*\thisrow{vv}+ \x},y = y/h,col sep=comma,unbounded coords=jump]{./figs/data/periodichill/fk03_finep1_x\x.csv};
        		\addlegendentry{$f_k = 0.3$}
    		
            \foreach \x in {1,...,8}{
                \addplot[color=black, style={ultra thin}, only marks, mark=o, mark options={scale=1.0}, mark repeat = 3, mark phase = 0]
    				table[x expr={10*\thisrow{vv}+ \x},y = y/h,col sep=comma,unbounded coords=jump]{./figs/data/periodichill/ref_x\x.csv};
    				
    			\addplot[color={Set1-A}, style={very thick}]
    				table[x expr={10*\thisrow{vv}+ \x},y = y/h,col sep=comma,unbounded coords=jump]{./figs/data/periodichill/LES_finep1_x\x.csv};

    		
    			\addplot[color={Set1-C}, style={very thick}]
    				table[x expr={10*\thisrow{vv}+ \x},y = y/h,col sep=comma,unbounded coords=jump]{./figs/data/periodichill/fk02_finep1_x\x.csv};
    		
    			\addplot[color={Set1-D}, style={very thick}]
    				table[x expr={10*\thisrow{vv}+ \x},y = y/h,col sep=comma,unbounded coords=jump]{./figs/data/periodichill/fk03_finep1_x\x.csv};
            }
    		
			\addplot[color=black, style={thin}, dashed]
			    table[x=x,y=y,col sep=comma,unbounded coords=jump]{./figs/data/periodichill/ph_geopoints.csv};
			\addplot[color=black, style={thin}, dashed]
			    table[x=x,y=y,col sep=comma,unbounded coords=jump]{./figs/data/periodichill/ph_geopoints.csv};
			    
    		\draw [dotted] (0,1) -- (0,3);
    		\draw [dotted] (1,.45) -- (1,3);
    		\draw [dotted] (2,0) -- (2,3);
    		\draw [dotted] (3,0) -- (3,3);
    		\draw [dotted] (4,0) -- (4,3);
    		\draw [dotted] (5,0) -- (5,3);
    		\draw [dotted] (6,0) -- (6,3);
    		\draw [dotted] (7,0) -- (7,3);
    		\draw [dotted] (8,.45) -- (8,3);
		\end{axis}

	\end{tikzpicture}}}
        \newline
        
        \caption{\label{fig:ph_vv}Time and span-averaged normal velocity variance profiles using a $\mathbb{P}_3$ (top) and $\mathbb{P}_1$ FR scheme (bottom). Profiles are scaled by a factor of 10 and shifted by 0, +1, +2, ..., +8, respectively, along the abscissa.}
    \end{figure}
        
    
    \begin{figure}[tbhp]
        \centering
        \subfloat[$\mathbb{P}_3$]{\label{fig:ph_uv_h}
        \adjustbox{width=0.7\linewidth,valign=b}{     \begin{tikzpicture}[spy using outlines={rectangle, height=3cm,width=2.3cm, magnification=3, connect spies}]
		\begin{axis}[name=plot1,
		    axis line style={latex-latex},
		    axis x line=left,
            axis y line=left,
            axis equal image,
            clip mode=individual,
		    xlabel={$u'v'/U_\infty^2$},
		    xtick={0, 1, 2, 3, 4, 5, 6, 7, 8},
    		xmin=-0.5,
    		xmax=9.2,
    		x tick label style={
        		/pgf/number format/.cd,
            	fixed,
            	precision=1,
        	    /tikz/.cd},
    		ylabel={$y/H$},
    		ytick={0, 1, 2, 3},
    		ymin=0,
    		ymax=3,
    		y tick label style={
        		/pgf/number format/.cd,
            	fixed,
            	precision=1,
        	    /tikz/.cd},
    		legend style={at={(1.0,0.5)},anchor=west},
    		legend cell align={left},
    		style={font=\small},
    		scale = 2]

            \def\x{0};
    			\addplot[color=black, style={ultra thin}, only marks, mark=o, mark options={scale=1.0}, mark repeat = 3, mark phase = 0]
    				table[x expr={15*\thisrow{uv}+ \x},y = y/h,col sep=comma,unbounded coords=jump]{./figs/data/periodichill/ref_x\x.csv};
    		    \addlegendentry{Ref. \citep{Frohlich2005}}
    		    
    			\addplot[color={Set1-A}, style={very thick}]
    				table[x expr={15*\thisrow{uv}+ \x},y = y/h,col sep=comma,unbounded coords=jump]{./figs/data/periodichill/LES_medp3_x\x.csv};
        		\addlegendentry{URDNS}

    		
    			\addplot[color={Set1-C}, style={very thick}]
    				table[x expr={15*\thisrow{uv}+ \x},y = y/h,col sep=comma,unbounded coords=jump]{./figs/data/periodichill/fk02_medp3_x\x.csv};
        		\addlegendentry{$f_k = 0.2$}
    		
    			\addplot[color={Set1-D}, style={ very thick}]
    				table[x expr={15*\thisrow{uv}+ \x},y = y/h,col sep=comma,unbounded coords=jump]{./figs/data/periodichill/fk03_medp3_x\x.csv};
        		\addlegendentry{$f_k = 0.3$}
    		
            \foreach \x in {1,...,8}{
                \addplot[color=black, style={ultra thin}, only marks, mark=o, mark options={scale=1.0}, mark repeat = 3, mark phase = 0]
    				table[x expr={15*\thisrow{uv}+ \x},y = y/h,col sep=comma,unbounded coords=jump]{./figs/data/periodichill/ref_x\x.csv};
    				
    			\addplot[color={Set1-A}, style={very thick}]
    				table[x expr={15*\thisrow{uv}+ \x},y = y/h,col sep=comma,unbounded coords=jump]{./figs/data/periodichill/LES_medp3_x\x.csv};

    		
    			\addplot[color={Set1-C}, style={very thick}]
    				table[x expr={15*\thisrow{uv}+ \x},y = y/h,col sep=comma,unbounded coords=jump]{./figs/data/periodichill/fk02_medp3_x\x.csv};
    		
    			\addplot[color={Set1-D}, style={very thick}]
    				table[x expr={15*\thisrow{uv}+ \x},y = y/h,col sep=comma,unbounded coords=jump]{./figs/data/periodichill/fk03_medp3_x\x.csv};
            }
    		
			\addplot[color=black, style={thin}, dashed]
			    table[x=x,y=y,col sep=comma,unbounded coords=jump]{./figs/data/periodichill/ph_geopoints.csv};
			\addplot[color=black, style={thin}, dashed]
			    table[x=x,y=y,col sep=comma,unbounded coords=jump]{./figs/data/periodichill/ph_geopoints.csv};
			    
    		\draw [dotted] (0,1) -- (0,3);
    		\draw [dotted] (1,.45) -- (1,3);
    		\draw [dotted] (2,0) -- (2,3);
    		\draw [dotted] (3,0) -- (3,3);
    		\draw [dotted] (4,0) -- (4,3);
    		\draw [dotted] (5,0) -- (5,3);
    		\draw [dotted] (6,0) -- (6,3);
    		\draw [dotted] (7,0) -- (7,3);
    		\draw [dotted] (8,.45) -- (8,3);
		\end{axis}

	\end{tikzpicture}}}
        \newline
        \subfloat[$\mathbb{P}_1$]{\label{fig:ph_uv_l}
        \adjustbox{width=0.7\linewidth,valign=b}{     \begin{tikzpicture}[spy using outlines={rectangle, height=3cm,width=2.3cm, magnification=3, connect spies}]
		\begin{axis}[name=plot1,
		    axis line style={latex-latex},
		    axis x line=left,
            axis y line=left,
            axis equal image,
            clip mode=individual,
		    xlabel={$u'v'/U_\infty^2$},
		    xtick={0, 1, 2, 3, 4, 5, 6, 7, 8},
    		xmin=-0.5,
    		xmax=9.2,
    		x tick label style={
        		/pgf/number format/.cd,
            	fixed,
            	precision=1,
        	    /tikz/.cd},
    		ylabel={$y/H$},
    		ytick={0, 1, 2, 3},
    		ymin=0,
    		ymax=3,
    		y tick label style={
        		/pgf/number format/.cd,
            	fixed,
            	precision=1,
        	    /tikz/.cd},
    		legend style={at={(1.0,0.5)},anchor=west},
    		legend cell align={left},
    		style={font=\small},
    		scale = 2]

            \def\x{0};
    			\addplot[color=black, style={ultra thin}, only marks, mark=o, mark options={scale=1.0}, mark repeat = 3, mark phase = 0]
    				table[x expr={15*\thisrow{uv}+ \x},y = y/h,col sep=comma,unbounded coords=jump]{./figs/data/periodichill/ref_x\x.csv};
    		    \addlegendentry{Ref. \citep{Frohlich2005}}
    		    
    			\addplot[color={Set1-A}, style={very thick}]
    				table[x expr={15*\thisrow{uv}+ \x},y = y/h,col sep=comma,unbounded coords=jump]{./figs/data/periodichill/LES_finep1_x\x.csv};
        		\addlegendentry{URDNS}

    		
    			\addplot[color={Set1-C}, style={very thick}]
    				table[x expr={15*\thisrow{uv}+ \x},y = y/h,col sep=comma,unbounded coords=jump]{./figs/data/periodichill/fk02_finep1_x\x.csv};
        		\addlegendentry{$f_k = 0.2$}
    		
    			\addplot[color={Set1-D}, style={ very thick}]
    				table[x expr={15*\thisrow{uv}+ \x},y = y/h,col sep=comma,unbounded coords=jump]{./figs/data/periodichill/fk03_finep1_x\x.csv};
        		\addlegendentry{$f_k = 0.3$}
    		
            \foreach \x in {1,...,8}{
                \addplot[color=black, style={ultra thin}, only marks, mark=o, mark options={scale=1.0}, mark repeat = 3, mark phase = 0]
    				table[x expr={15*\thisrow{uv}+ \x},y = y/h,col sep=comma,unbounded coords=jump]{./figs/data/periodichill/ref_x\x.csv};
    				
    			\addplot[color={Set1-A}, style={very thick}]
    				table[x expr={15*\thisrow{uv}+ \x},y = y/h,col sep=comma,unbounded coords=jump]{./figs/data/periodichill/LES_finep1_x\x.csv};

    		
    			\addplot[color={Set1-C}, style={very thick}]
    				table[x expr={15*\thisrow{uv}+ \x},y = y/h,col sep=comma,unbounded coords=jump]{./figs/data/periodichill/fk02_finep1_x\x.csv};
    		
    			\addplot[color={Set1-D}, style={very thick}]
    				table[x expr={15*\thisrow{uv}+ \x},y = y/h,col sep=comma,unbounded coords=jump]{./figs/data/periodichill/fk03_finep1_x\x.csv};
            }
    		
			\addplot[color=black, style={thin}, dashed]
			    table[x=x,y=y,col sep=comma,unbounded coords=jump]{./figs/data/periodichill/ph_geopoints.csv};
			\addplot[color=black, style={thin}, dashed]
			    table[x=x,y=y,col sep=comma,unbounded coords=jump]{./figs/data/periodichill/ph_geopoints.csv};
			    
    		\draw [dotted] (0,1) -- (0,3);
    		\draw [dotted] (1,.45) -- (1,3);
    		\draw [dotted] (2,0) -- (2,3);
    		\draw [dotted] (3,0) -- (3,3);
    		\draw [dotted] (4,0) -- (4,3);
    		\draw [dotted] (5,0) -- (5,3);
    		\draw [dotted] (6,0) -- (6,3);
    		\draw [dotted] (7,0) -- (7,3);
    		\draw [dotted] (8,.45) -- (8,3);
		\end{axis}

	\end{tikzpicture}}}
        \newline
        
        \caption{\label{fig:ph_uv}Time and span-averaged streamwise-normal velocity covariance profiles using a $\mathbb{P}_3$ (top) and $\mathbb{P}_1$ FR scheme (bottom). Profiles are scaled by a factor of 10 and shifted by 0, +1, +2, ..., +8, respectively, along the abscissa.}
    \end{figure}
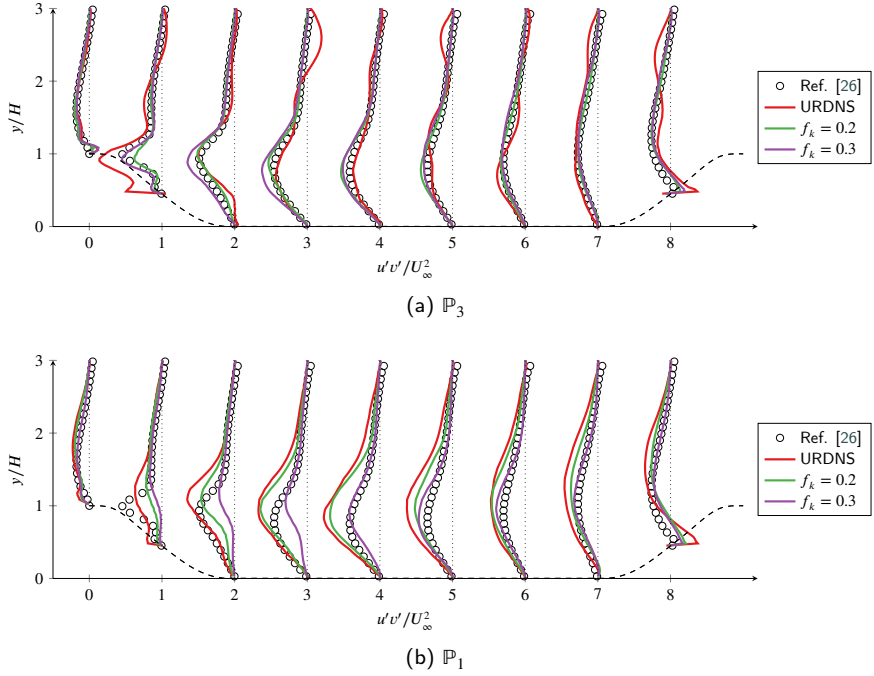
        

The profiles of the normal velocity variance are shown in \cref{fig:ph_vv}. For the low-order method, poor agreement with the reference data was generally observed for both PANS and URDNS, with the PANS approach underpredicting the variance in the separation region and the URDNS approach overpredicting the variance outside of the separation region. However, outside of the separation region, the prediction by the low-order PANS approach significantly improved. When switching to the high-order method, the results of the URDNS approach degraded, with large oscillations observed in the variance profiles. This effect can likely be attributed to aliasing-driven instabilities evident in high-order approximations of under-resolved turbulent flows, which can introduce spurious high-frequency oscillations in the flow that are more evident in higher-order statistics \citep{Cox2021}. When using the high-order method for the PANS approach, this effect was not seen, likely due to the suppression of aliasing errors as a result of the introduction of a physically appropriate eddy viscosity. Furthermore, significant improvements were observed in the variance profiles between the low-order and high-order PANS approaches, with the high-order PANS profiles showing excellent agreement with the reference data for both $f_k$ values and notably less deviation in the results between different $f_k$ values.

Similar observations were seen in the streamwise-normal velocity covariance profiles, shown in \cref{fig:ph_uv}. No appreciable improvement in the covariance profiles was observed when switching from low-order to high-order URDNS, with the high-order approach showing better results in the separation region but at the expense of spurious oscillations outside of the separation region as in \cref{fig:ph_vv}. The low-order PANS approach showed reasonable results outside of the separation region with $f_k = 0.3$, but underpredicted the magnitude of the covariance aft of the separation point for both values of $f_k$. The high-order PANS approach showed good agreement with the reference data for both $f_k$ values, with significantly better prediction within the separation region and monotonic covariance profiles outside of the separation region. The decrease in sensitivity to the $f_k$ parameter when switching to the high-order approach was not as evident for the streamwise-normal velocity covariance as it was with the streamwise and normal variance.

\subsection{Cylinder}
Due to the variety of physical phenomena that appear, the flow around a circular cylinder is of interest for many applications in fluid dynamics. At moderately-low Reynolds numbers (400-5000), the flow lies in a subcritical regime where the transition of the laminar shear layer to a turbulent wake presents a challenge for numerical studies as the flow physics are very sensitive to the methods used which can lead to significant discrepancies in the results between various approaches \citep{Parnaudeau2008}. The flow at a Reynolds number ($Re_D$) of 3900, based on the freestream velocity $U_\infty$ and cylinder diameter $D$, lies in this regime, and as such, there exists vast amounts of numerical and experimental data for this configuration \citep{Lehmkuhl2013, Parnaudeau2008, Witherden2015, Ong1996, Kravchenko2000, Ma2000}. This problem was explored to conduct a comparison between low-order and high-order PANS and URDNS methods for resolving more complex flow physics. These results were compared to experimental and numerical data, and a DNS study was performed to provide additional data for comparison that was not available in the literature. The comparisons were carried out at levels of resolution which can be considered to be "moderately-resolved", where the prediction of first-order statistics can be done to a reasonable accuracy without turbulence modeling (i.e., implicit LES/under-resolved DNS). As such, the metric for comparison between the methods was their ability to resolve the more complex flow physics of the problem, such as the vortex shedding and Kelvin--Helmholtz frequencies and the presence of coherent turbulent structures in the wake.

\begin{figure}[tbhp]
    \centering
    \subfloat[Geometry]{
    \adjustbox{width=0.47\linewidth,valign=b}{     \begin{tikzpicture}[spy using outlines={rectangle, height=3cm,width=2.3cm, magnification=3, connect spies}]
		\begin{axis}[name=plot1,
		    axis line style={draw=none},
		    tick style={draw=none},
		    axis x line=left,
            axis y line=left,
            axis equal image,
            clip mode=individual,
    		xmin=-10,
    		xmax=25,
    		xticklabels={,,},
    		ymin=-10,
    		ymax=10,
    		yticklabels={,,},
    		style={font=\small},
    		scale = 1]

            \draw[-,fill=gray] (axis cs:0.0, 0.0) circle[radius=0.5];
            
		    \draw[-] (axis cs:0.0, 10.0) -- (axis cs:25.0, 10.0);
		    \draw[-] (axis cs:0.0, -10.0) -- (axis cs:25.0, -10.0);
		    \draw[-] (axis cs:25.0, 10.0) -- (axis cs:25.0, -10.0);
		    \draw[-] (0,10) arc (90:270:10);

		    \draw[<->] (axis cs:0, -11) -- (axis cs:25.0, -11);
		    \node at (axis cs:12.5,-12) {$25D$};
		    
		    \draw[<->] (axis cs:26, -10) -- (axis cs:26, 10);
		    \node at (axis cs:28.5,0) {$10D$};

		    \draw[<->] (axis cs:0, -11) -- (axis cs:25.0, -11);
		    
		    
			    
		\end{axis}

	\end{tikzpicture}}}
    \subfloat[Mesh]{
    \adjustbox{width=0.53\linewidth,valign=b}{\includegraphics[]{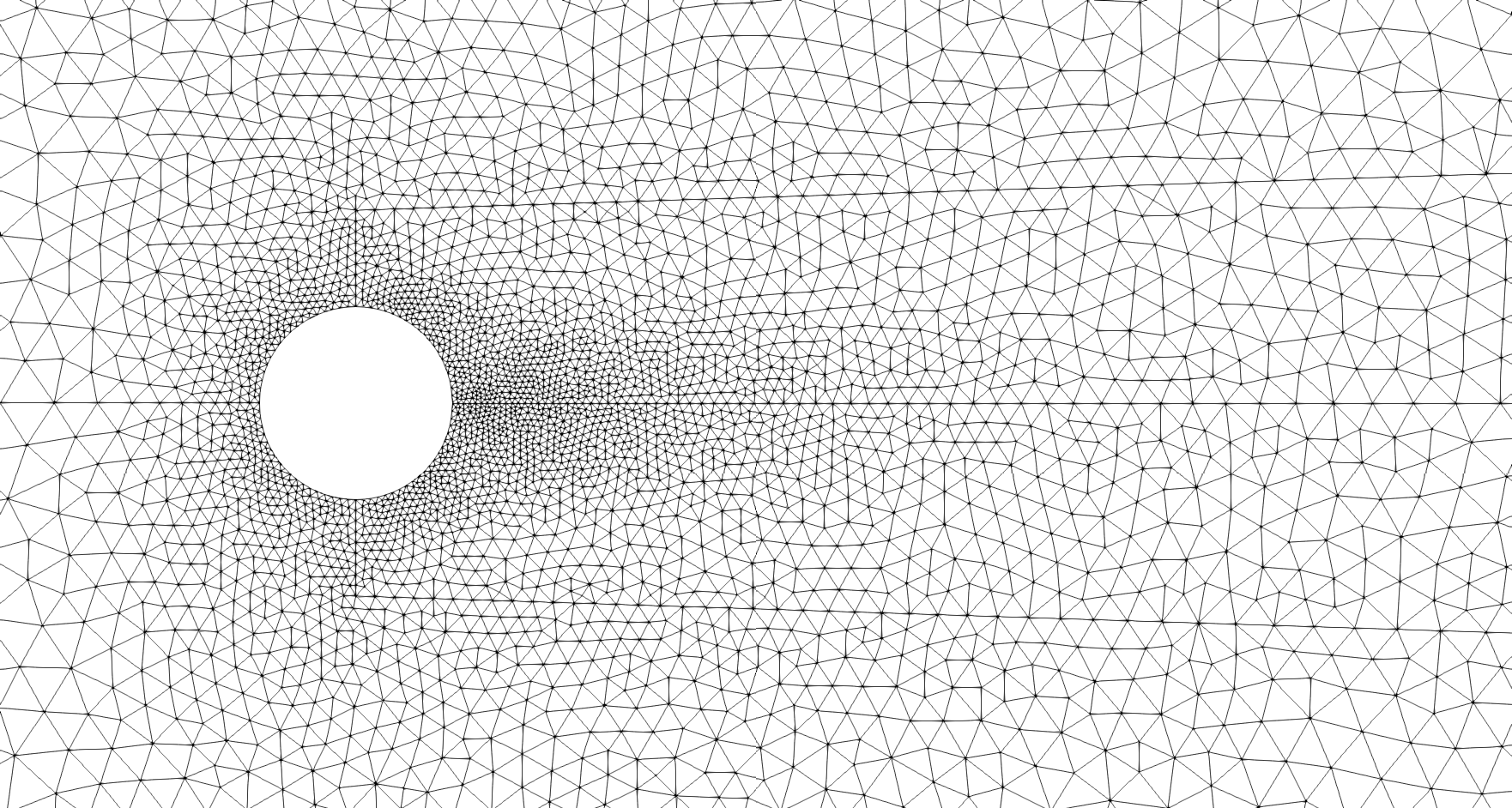}}}
    \newline
    \caption{\label{fig:cyl_geo} Cross-section of the cylinder geometry in the streamwise-spanwise plane (left) and mesh for the $\mathbb{P}_3$ approach in the cylinder wake region (right). }
\end{figure}

The simulations were performed using a C-grid domain on $[x/D, y/D, z/D] \in [-10, 25] \times [-10, 10] \times [0, \pi]$, where $x$, $y$, and $z$ denote the streamwise, normal, and spanwise directions, respectively. The domain and grid are shown in \cref{fig:cyl_geo}. At the cylinder surface, a no-slip, adiabatic wall boundary condition was applied, with the turbulence variables taking on the values in \cref{eq:pansbcs}. Periodic boundary conditions were applied along the spanwise direction. At the inlet, a uniform density and momentum were fixed, and the turbulence variables were set according to \citet{Menter1994} as
\begin{equation}
    k_u = 10^{-3} f_k \frac{U_\infty^2}{Re_D}, \quad \quad \omega_u = f_\omega \frac{U_\infty}{D}.
\end{equation}
At the outlet, the pressure was set to a fixed value corresponding to a Mach number of 0.1 while the remaining variables were free. For all simulations, the flow was assumed to be fully developed after $t = D/U_\infty = 100$, after which statistical quantities were gathered and analyzed until $t = D/U_\infty = 300$. 

Three unstructured, prismatic grids were generated by extruding a 2D, triangular grid along the spanwise axis, one for the low-order ($\mathbb{P}_1$) PANS/URDNS approaches, one for the high-order ($\mathbb{P}_3$) PANS/URDNS approaches, and one for the DNS. A description of the characteristics of these meshes is shown in \cref{tab:cyl_grids}, where $N$ denotes the number of elements and $DOF$ denotes the degrees of freedom. The values corresponding to the unstructured, triangular elements in the streamwise-normal plane are denoted by the subscript $xy$ while the values corresponding to the structured extrusion along the spanwise direction are denoted by the subscript $z$. The low-order mesh was generated by uniformly sub-dividing the high-order mesh along each direction. As a result of the subdivision of the triangular surfaces, the low-order mesh had approximately 20\% more degrees of freedom than the high-order mesh. In comparison to the low-order PANS approach used for this configuration in \citet{Pereira2018}, the grids in the present work consist of approximately 20-30\% fewer degrees of freedom.

\begin{figure}[tbhp]
    \centering
    \begin{tabular}{r c c c c}\toprule
        Grid & Method & Order & $N_{xy} \times N_z$ & $DOF_{xy} \times DOF_z$   \\ \midrule
        
        B1 &
        PANS/URDNS &
        $\mathbb{P}_1$ &
        $26,000 \times 20$ &
        $\hphantom{0}78,000 \times 40$  \\
        
        B2 &  
        PANS/URDNS &
        $\mathbb{P}_3$ &
        $\hphantom{0}6,500 \times 10$ &
        $\hphantom{0}65,000 \times 40$ \\ 
        
        B3 &  
        DNS &
        $\mathbb{P}_3$ &
        $ 58,000 \times 24$ &
        $ 580,000 \times 96$ \\ 
        \bottomrule
    \end{tabular}
    
    \captionof{table}{\label{tab:cyl_grids}Description of mesh characteristics for the cylinder case. }
\end{figure}

The DNS study was performed using a $\mathbb{P}_3$ approximation with roughly 55.6 million degrees of freedom and an identical problem configuration (excluding the turbulence model). Through \emph{a posteriori} analysis, the minimum Kolmogorov length scale in the domain was found to be $\eta/D = 0.011$. Therefore, this resolution was sufficient to achieve a $\Delta s_{xy}/\eta$ ratio between 0.25 and 0.83 for $x/D < 10$, where $\Delta s_{xy}$ is the average subcell size within an element in the streamwise-normal plane. Furthermore, the $\Delta s_{z}/\eta$ ratio was 0.85, where $\Delta s_{z}$ is the average subcell size within an element in the spanwise direction. The results of the DNS study were used to perform a Proper Orthogonal Decomposition (POD) analysis to identify the presence of coherent turbulent structures which may not be evident in the first and second-order statistics. This method, described in detail in \cref{app:pod}, was used to characterize the time-dependent velocity field by a set of orthonormal spatial modes, such that the modes containing the most energy corresponded to the most dominant features in the flow field. To evaluate the ability of the various approaches in this work in predicting the dominant flow physics of the problem, the highest energy mode of the streamwise velocity fluctuations was compared to the DNS results.

After averaging across the time-averaging horizon and along the spanwise direction, the PANS and URDNS approaches were compared to the DNS results of the present work, the LES results of \citet{Parnaudeau2008}, and the LES results of \citet{Witherden2015}. For comparison with the work of \citet{Witherden2015} where individual data is presented for the two distinct shedding modes across a large time horizon, we take the average of the two modes. These results were analyzed with respect to the first and second-order statistics, frequency spectra, and POD modes.

\begin{figure}[tbhp]
    \centering
        \subfloat[$\mathbb{P}_3$]{\label{fig:cyl_center_HO} \adjustbox{width=0.4\linewidth,valign=b}{     \begin{tikzpicture}[spy using outlines={rectangle, height=3cm,width=2.3cm, magnification=3, connect spies}]
		\begin{axis}[name=plot1,
		    axis line style={latex-latex},
		    axis x line=left,
            axis y line=left,
            clip mode=individual,
		    xlabel={$x/D$},
		    xtick={0.5, 1.0, 2, 3, 4, 5},
    		xmin=0.5,
    		xmax=5,
    		x tick label style={
        		/pgf/number format/.cd,
            	fixed,
            	precision=1,
        	    /tikz/.cd},
    		ylabel={$u/U_\infty$},
    		ytick={-1, -0.5, 0, 0.5, 1.0},
    		ymin=-0.5,
    		ymax=1,
    		y tick label style={
        		/pgf/number format/.cd,
            	fixed,
            	precision=1,
        	    /tikz/.cd},
    		legend style={at={(1.0,0.03)},anchor=south east,font=\small},
    		legend cell align={left},
    		style={font=\normalsize}]
    		
    		
			\addplot[color=black, style={ultra thin}, only marks, mark=o, mark options={scale=1.0}, mark repeat = 20, mark phase = 0]
				table[x =x,y =u,col sep=comma,unbounded coords=jump]{./figs/data/cyl_DNS/cylinder_DNS_centerline.csv};
    		\addlegendentry{DNS}
    		
			\addplot[color=black, style={ultra thin}, only marks, mark=diamond, mark options={scale=1.0}, mark repeat = 2, mark phase = 0]
				table[x index=0,y index=1,col sep=comma,unbounded coords=jump]{./figs/data/cyl_ref_exp/parn_u_centerline.csv};
    		\addlegendentry{Ref. \citep{Parnaudeau2008}}
    		
			\addplot[color=black, style={ultra thin}, only marks, mark=*, mark options={scale=0.5}, mark repeat = 3, mark phase = 0]
				table[x index=0,y index=1,col sep=comma,unbounded coords=jump]{./figs/data/cyl_ref_DNS_M/modem_u_centerline.csv};
    		\addlegendentry{Ref. \citep{Witherden2015}}

			\addplot[color={Set1-A}, style={very thick}]
				table[x=x,y=u,col sep=comma,unbounded coords=jump]{./figs/data/cyl_LES/cylinderLES_medp3_centerline.csv};
    		\addlegendentry{URDNS}
    		
			\addplot[color={Set1-B}, style={very thick}]
				table[x =Points:0,y=Avg_U_average,col sep=comma,unbounded coords=jump]{./figs/data/cyl_fk01/cylinderPANS_medp3_centerline.csv};
    		\addlegendentry{$f_k = 0.1$}
    		
			\addplot[color={Set1-C}, style={very thick}]
				table[x =Points:0,y=Avg_U_average,col sep=comma,unbounded coords=jump]{./figs/data/cyl_fk02/cylinderPANS_medp3_centerline.csv};
    		\addlegendentry{$f_k = 0.2$}
    		
			\addplot[color={Set1-D}, style={very thick}]
				table[x =Points:0,y=Avg_U_average,col sep=comma,unbounded coords=jump]{./figs/data/cyl_fk03/cylinderPANS_medp3_centerline.csv};
    		\addlegendentry{$f_k = 0.3$}

        \draw [dotted] (0.5,0) -- (5,0);
		\end{axis}

	\end{tikzpicture}}}
    ~
        \subfloat[$\mathbb{P}_1$]{\label{fig:cyl_center_LO} \adjustbox{width=0.4\linewidth,valign=b}{\includegraphics[]{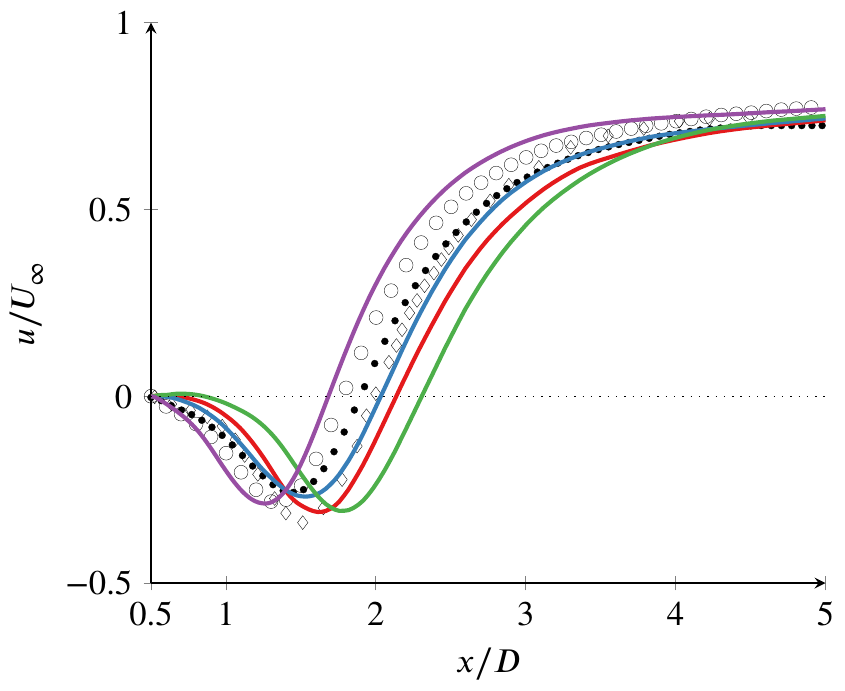}}}
    \caption{\label{fig:cyl_center}Time and span-averaged streamwise velocity profiles along the centerline ($y/D = 0$).}
\end{figure}

The profiles of the averaged centerline streamwise velocity are shown in \cref{fig:cyl_center} in comparison to the reference data. Between all of the approaches, reasonable approximations of the centerline streamwise velocity were obtained when taking into account the variation in the reference results. For the URDNS approaches, the low-order method overpredicted the edge of the recirculation region, but this was remedied with the high-order method. For the PANS approaches, the effects of the higher-order discretization were not immediately evident. With the low-order method, the efficacy of the PANS model was sensitive to the value of $f_k$, with $f_k = 0.1$ showing reasonable agreement with the LES of \citet{Parnaudeau2008} and $f_k = 0.3$ showing reasonable agreement with the DNS results. With the high-order method, less variation of the results with respect to $f_k$ was seen. Excellent agreement between the $f_k = 0.1$ results and the LES of \citet{Witherden2015} was observed. The introduction of the PANS model generally prolonged the recirculation region with the high-order method, whereas for the low-order method, the effect was not clear. 

\begin{figure}[tbhp!]
        \centering
        \subfloat[$x/D = 1.06$]{\label{fig:u_x1p06} \adjustbox{width=0.4\linewidth,valign=b}{     \begin{tikzpicture}[spy using outlines={rectangle, height=3cm,width=2.3cm, magnification=3, connect spies}]
		\begin{axis}[name=plot1,
		    axis line style={latex-latex},
		    axis x line=left,
            axis y line=left,
            clip mode=individual,
		    xlabel={$y/D$},
		    xtick={-2, -1, 0, 1, 2},
    		xmin=-2,
    		xmax=2,
    		x tick label style={
        		/pgf/number format/.cd,
            	fixed,
            	precision=1,
        	    /tikz/.cd},
    		ylabel={$u/U_\infty$},
    		ytick={-0.5, 0, 0.5, 1.0, 1.5},
    		ymin=-0.5,
    		ymax=1.5,
    		y tick label style={
        		/pgf/number format/.cd,
            	fixed,
            	precision=1,
        	    /tikz/.cd},
    		legend style={at={(1.0,0.03)},anchor=south east,font=\small},
    		legend cell align={left},
    		style={font=\normalsize}]
    		

			\addplot[color=black, style={ultra thin}, only marks, mark=o, mark options={scale=1.0}, mark repeat = 20, mark phase = 0]
				table[x=y,y=u,col sep=comma,unbounded coords=jump]{./figs/data/cyl_DNS/cylinder_DNS_x1p06.csv};
    		\addlegendentry{DNS}
    		
			\addplot[color=black, style={ultra thin}, only marks, mark=diamond, mark options={scale=1.0}, mark repeat = 2, mark phase = 0]
				table[x index=0,y index=1,col sep=comma,unbounded coords=jump]{./figs/data/cyl_ref_exp/parn_u_x1p06.csv};
    		\addlegendentry{Ref. \citep{Parnaudeau2008}}
    		
			\addplot[color=black, style={ultra thin},  only marks, mark=*, mark options={scale=0.5}, mark repeat = 3, mark phase = 0]
				table[x index=0,y index=1,col sep=comma,unbounded coords=jump]{./figs/data/cyl_ref_DNS_M/modem_u_x1p06.csv};
    		\addlegendentry{Ref. \citep{Witherden2015}}

			\addplot[color={Set1-A}, style={very thick}]
				table[x =y,y=u,col sep=comma,unbounded coords=jump]{./figs/data/cyl_LES/cylinderLES_medp3_x1p06.csv};
    		\addlegendentry{URDNS}

			\addplot[color={Set1-B}, style={very thick}]
				table[x =Points:1,y=Avg_U_average,col sep=comma,unbounded coords=jump]{./figs/data/cyl_fk01/cylinderPANS_medp3_x1p06.csv};
    		\addlegendentry{$f_k = 0.1$}
    		
			\addplot[color={Set1-C}, style={very thick}]
				table[x =Points:1,y=Avg_U_average,col sep=comma,unbounded coords=jump]{./figs/data/cyl_fk02/cylinderPANS_medp3_x1p06.csv};
    		\addlegendentry{$f_k = 0.2$}
    		
			\addplot[color={Set1-D}, style={very thick}]
				table[x =Points:1,y=Avg_U_average,col sep=comma,unbounded coords=jump]{./figs/data/cyl_fk03/cylinderPANS_medp3_x1p06.csv};
    		\addlegendentry{$f_k = 0.3$}

			\addplot[color={Set1-A}, style={very thick, dotted}]
				table[x =y,y=u,col sep=comma,unbounded coords=jump]{./figs/data/cyl_LES_LO/cylinderLES_finep1_x1p06.csv};
			\addplot[color={Set1-B}, style={very thick, dotted}]
				table[x =y,y=u,col sep=comma,unbounded coords=jump]{./figs/data/cyl_fk01_LO/cylinderPANS_finep1_x1p06.csv};
			\addplot[color={Set1-C}, style={very thick, dotted}]
				table[x =y,y=u,col sep=comma,unbounded coords=jump]{./figs/data/cyl_fk02_LO/cylinderPANS_finep1_x1p06.csv};
			\addplot[color={Set1-D}, style={very thick, dotted}]
				table[x =y,y=u,col sep=comma,unbounded coords=jump]{./figs/data/cyl_fk03_LO/cylinderPANS_finep1_x1p06.csv};

		\end{axis}

	\end{tikzpicture}}}
        ~
        \subfloat[$x/D = 1.06$]{\label{fig:v_x1p06} \adjustbox{width=0.4\linewidth,valign=b}{\includegraphics[]{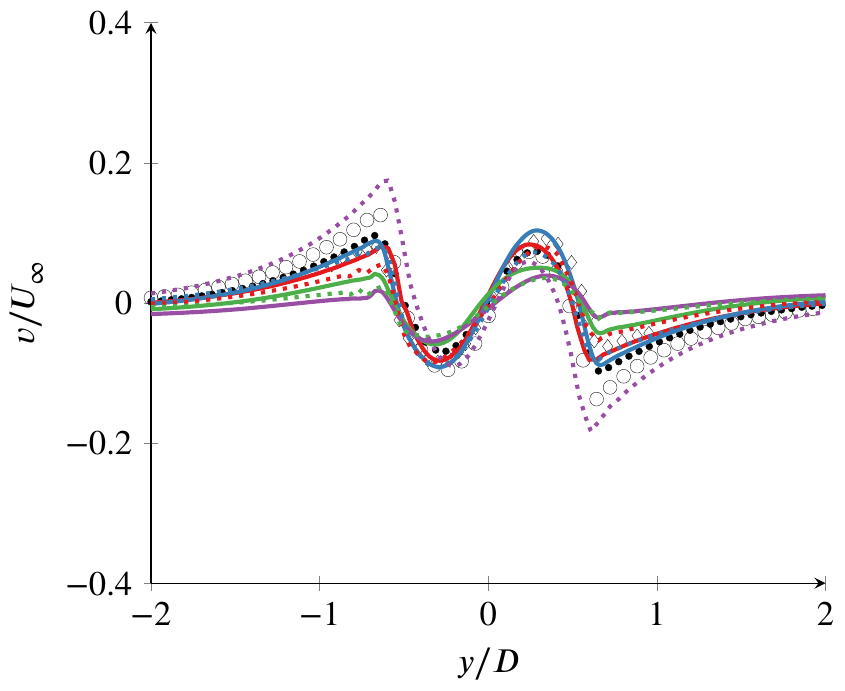}}}
        ~
        \newline
        \subfloat[$x/D = 1.54$]{\label{fig:u_x1p54} \adjustbox{width=0.4\linewidth,valign=b}{\includegraphics[]{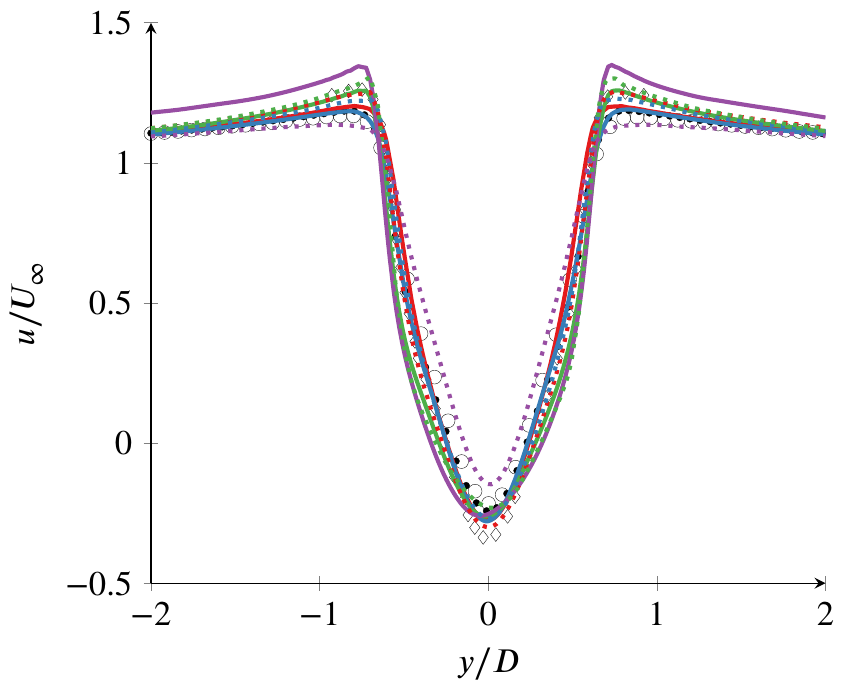}}}
        ~
        \subfloat[$x/D = 1.54$]{\label{fig:v_x1p54} \adjustbox{width=0.4\linewidth,valign=b}{\includegraphics[]{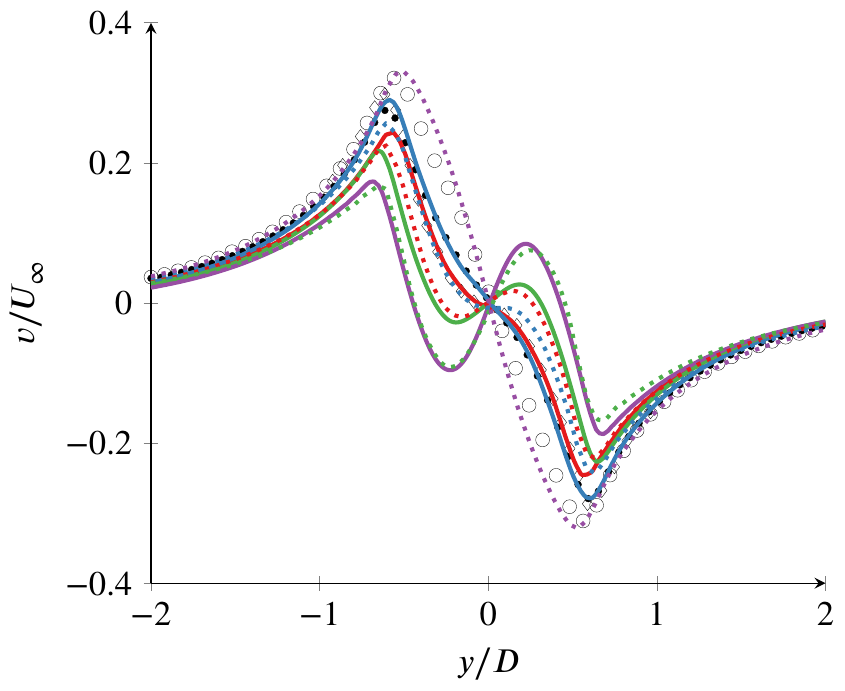}}}
        ~
        \newline
        \subfloat[$x/D = 2.02$]{\label{fig:u_x2p02} \adjustbox{width=0.4\linewidth,valign=b}{\includegraphics[]{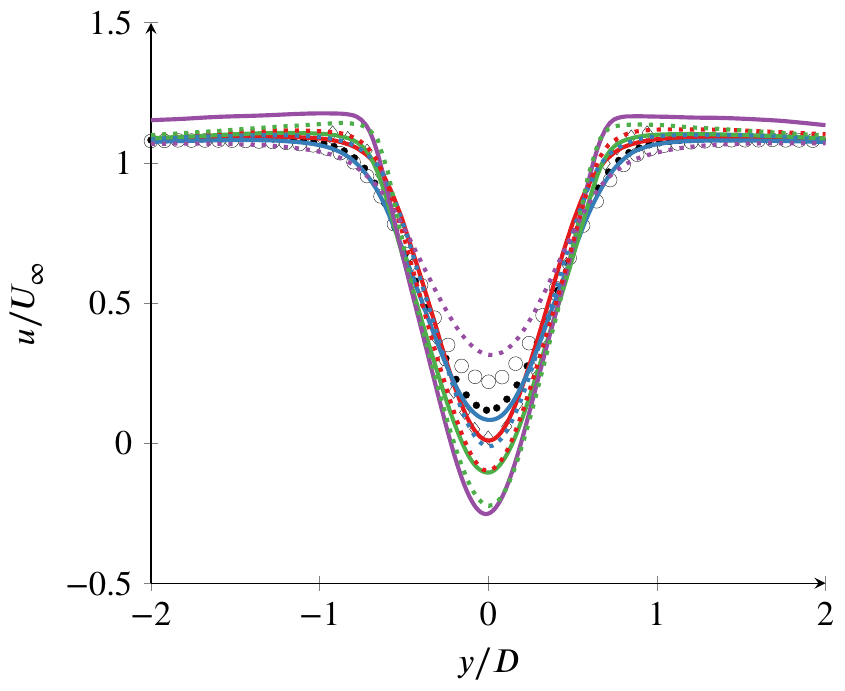}}}
        ~
        \subfloat[$x/D = 2.02$]{\label{fig:v_x2p02} \adjustbox{width=0.4\linewidth,valign=b}{\includegraphics[]{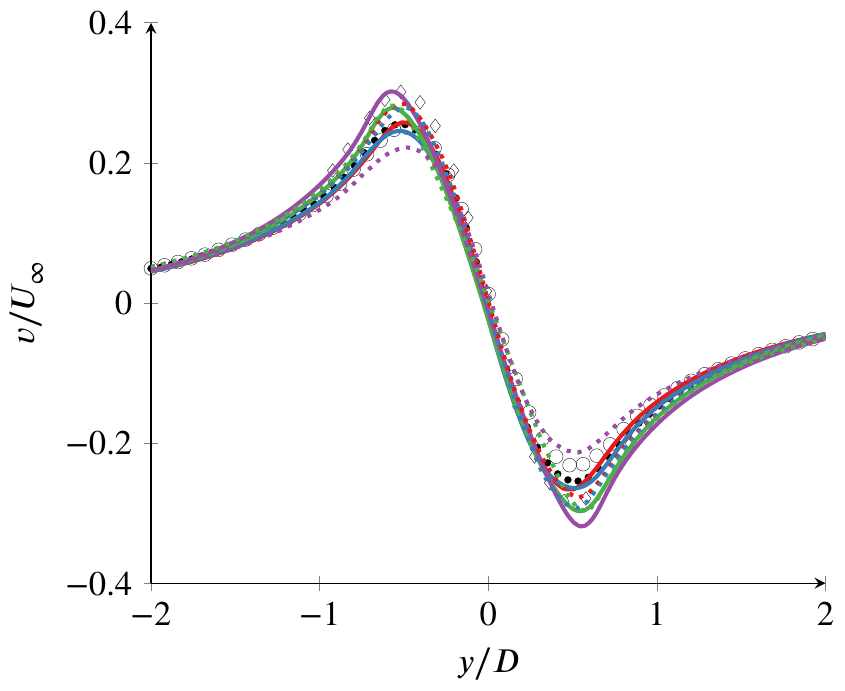}}}
        ~
        \newline
        \caption{\label{fig:cyl_xslices}Time and span-averaged streamwise (left) and normal (right) velocity profiles at $x/D = 1.06$, $1.54$, and $2.02$ using a $\mathbb{P}_3$ (solid line) and $\mathbb{P}_1$ FR scheme (dotted line).}
    \end{figure}

The averaged streamwise and normal velocity profiles at $x/D = 1.06$, $1.54$, and $2.02$ are shown in \cref{fig:cyl_xslices}. For the streamwise velocity profiles at $x/D = 1.06$ and $1.54$, minimal differences were observed between the low-order and high-order PANS and URDNS approaches, with all methods showing reasonable agreement with the reference data. At $x/D = 2.02$, effects similar to the centerline velocity profiles were seen, where the sensitivity of the PANS approach to the $f_k$ parameter decreased with the use of the high-order approach. For the normal velocity profiles, the differences between the various approaches were most evident at $x/D = 1.54$ but without clear distinction in the effects of higher-order discretizations on the PANS and URDNS approaches. Across the range of sampling locations, the best accuracy for the low-order approach was observed with $f_k = 0.3$, showing good agreement with the DNS results, and the best accuracy for the high-order approach was observed with $f_k = 0.1$, showing excellent agreement with the LES of \citet{Witherden2015}. This observation is consistent with the PANS methodology, as the increase in fidelity afforded by high-order methods coincides with a decrease in the unresolved portion of the flow. 

    
\begin{figure}[tbhp]
        \centering
        \subfloat[$x/D=1.06$]{\label{fig:uu_x1p06} \adjustbox{width=0.3\linewidth,valign=b}{     \begin{tikzpicture}[spy using outlines={rectangle, height=3cm,width=2.3cm, magnification=3, connect spies}]
		\begin{axis}[name=plot1,
		    axis line style={latex-latex},
		    axis x line=left,
            axis y line=left,
            clip mode=individual,
		    xlabel={$y/D$},
		    xtick={-1, -0.5, 0, 0.5, 1},
    		xmin=-1,
    		xmax=1,
    		x tick label style={
        		/pgf/number format/.cd,
            	fixed,
            	precision=1,
        	    /tikz/.cd},
    		ylabel={$u'u'/U_\infty^2$},
    		ytick={0, 0.1, 0.2, 0.3, 0.4, 0.5},
    		ymin=-0.02,
    		ymax=0.5,
    		y tick label style={
        		/pgf/number format/.cd,
            	fixed,
            	precision=1,
        	    /tikz/.cd},
    		legend style={at={(0.95,1.0)},anchor=north east, font=\small},
    		legend cell align={left},
    		style={font=\normalsize},
    		scale = 0.7,
    		legend columns=3,
            transpose legend]
    		

			\addplot[color=black, style={ultra thin}, only marks, mark=o, mark options={scale=1.0}, mark repeat = 20, mark phase = 0]
				table[x=y,y=uu,col sep=comma,unbounded coords=jump]{./figs/data/cyl_DNS/cylinder_DNS_x1p06.csv};
    		\addlegendentry{DNS}
				
			\addplot[color=black, style={ultra thin}, only marks, mark=diamond, mark options={scale=1.0}, mark repeat = 2, mark phase = 0]
				table[x index=0,y index=1,col sep=comma,unbounded coords=jump]{./figs/data/cyl_ref_exp/parn_uu_x1p06.csv};
    		\addlegendentry{Ref. \citep{Parnaudeau2008}}

			\addplot[color={Set1-A}, style={very thick}]
				table[x=y,y=uu,col sep=comma,unbounded coords=jump]{./figs/data/cyl_LES/cylinderLES_medp3_x1p06.csv};
    		\addlegendentry{URDNS}

			\addplot[color={Set1-B}, style={very thick}]
				table[x=y,y=uu,col sep=comma,unbounded coords=jump]{./figs/data/cyl_fk01/cylinderPANS_medp3_fluc_x1p06.csv};
    		\addlegendentry{$f_k = 0.1$}
    		
			\addplot[color={Set1-C}, style={very thick}]
				table[x=y,y=uu,col sep=comma,unbounded coords=jump]{./figs/data/cyl_fk02/cylinderPANS_medp3_fluc_x1p06.csv};
    		\addlegendentry{$f_k = 0.2$}
    		
			\addplot[color={Set1-D}, style={very thick}]
				table[x=y,y=uu,col sep=comma,unbounded coords=jump]{./figs/data/cyl_fk03/cylinderPANS_medp3_fluc_x1p06.csv};
    		\addlegendentry{$f_k = 0.3$}
    		
			\addplot[color={Set1-A}, style={very thick, dotted}]
				table[x =y,y=uu,col sep=comma,unbounded coords=jump]{./figs/data/cyl_LES_LO/cylinderLES_finep1_x1p06.csv};
			\addplot[color={Set1-B}, style={very thick, dotted}]
				table[x =y,y=uu,col sep=comma,unbounded coords=jump]{./figs/data/cyl_fk01_LO/cylinderPANS_finep1_x1p06.csv};
			\addplot[color={Set1-C}, style={very thick, dotted}]
				table[x =y,y=uu,col sep=comma,unbounded coords=jump]{./figs/data/cyl_fk02_LO/cylinderPANS_finep1_x1p06.csv};
			\addplot[color={Set1-D}, style={very thick, dotted}]
				table[x =y,y=uu,col sep=comma,unbounded coords=jump]{./figs/data/cyl_fk03_LO/cylinderPANS_finep1_x1p06.csv};
    		
		\end{axis}

	\end{tikzpicture}}}
        ~
        \subfloat[$x/D=1.06$]{\label{fig:vv_x1p06} \adjustbox{width=0.3\linewidth,valign=b}{\includegraphics[]{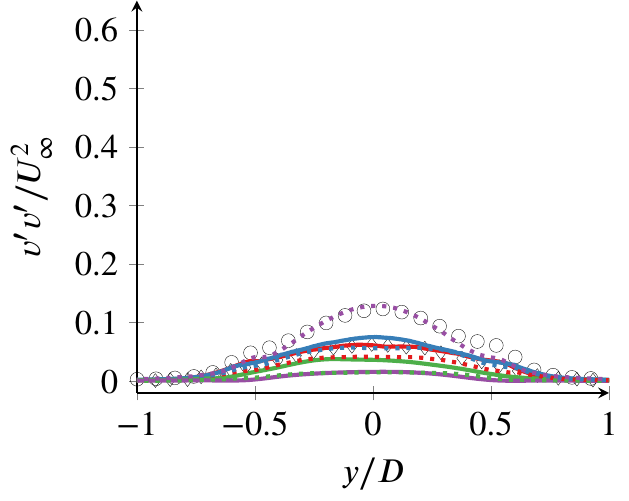}}}
        ~
        \subfloat[$x/D=1.06$]{\label{fig:uv_x1p06} \adjustbox{width=0.3\linewidth,valign=b}{\includegraphics[]{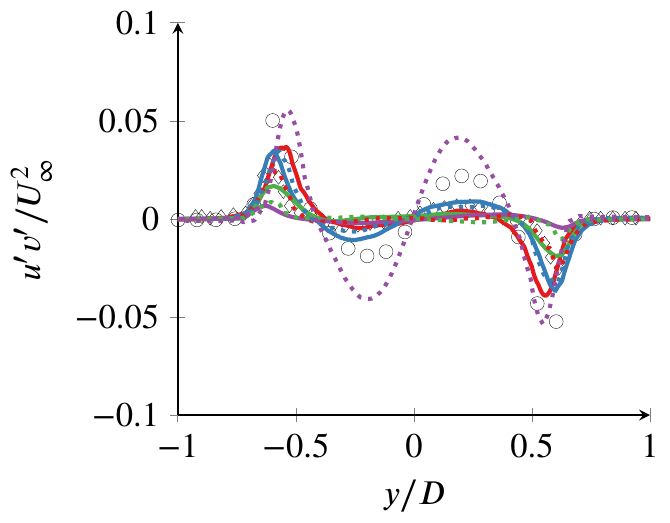}}}
        ~
        \newline
        \subfloat[$x/D=1.54$]{\label{fig:uu_x1p54} \adjustbox{width=0.3\linewidth,valign=b}{\includegraphics[]{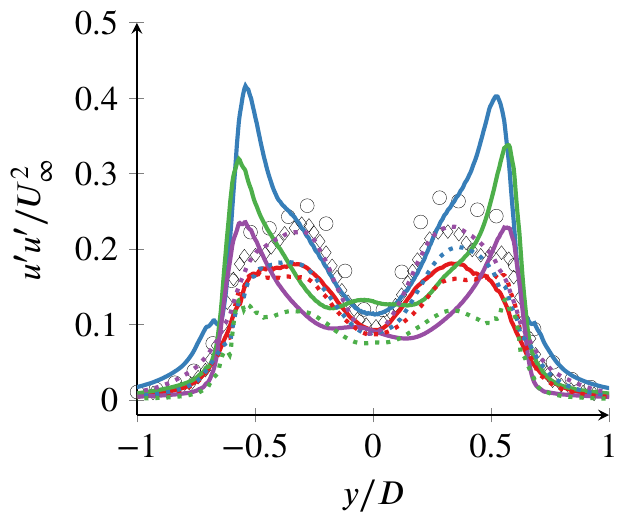}}}
        ~
        \subfloat[$x/D=1.54$]{\label{fig:vv_x1p54} \adjustbox{width=0.3\linewidth,valign=b}{\includegraphics[]{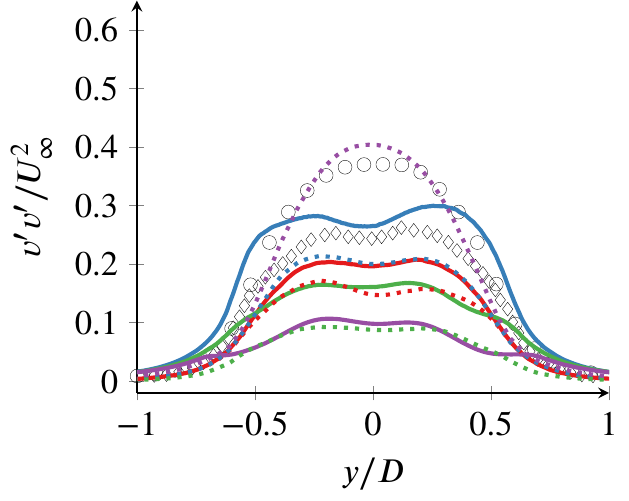}}}
        ~
        \subfloat[$x/D=1.54$]{\label{fig:uv_x1p54} \adjustbox{width=0.3\linewidth,valign=b}{\includegraphics[]{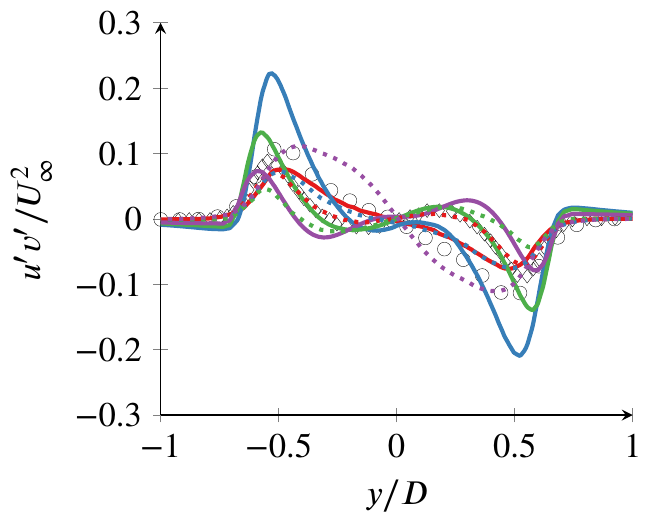}}}
        ~
        \newline
        \subfloat[$x/D=2.02$]{\label{fig:uu_x2p02} \adjustbox{width=0.3\linewidth,valign=b}{\includegraphics[]{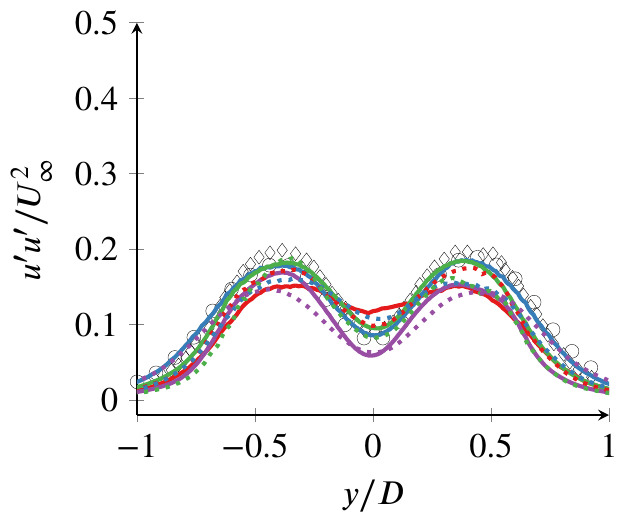}}}
        ~
        \subfloat[$x/D=2.02$]{\label{fig:vv_x2p02} \adjustbox{width=0.3\linewidth,valign=b}{\includegraphics[]{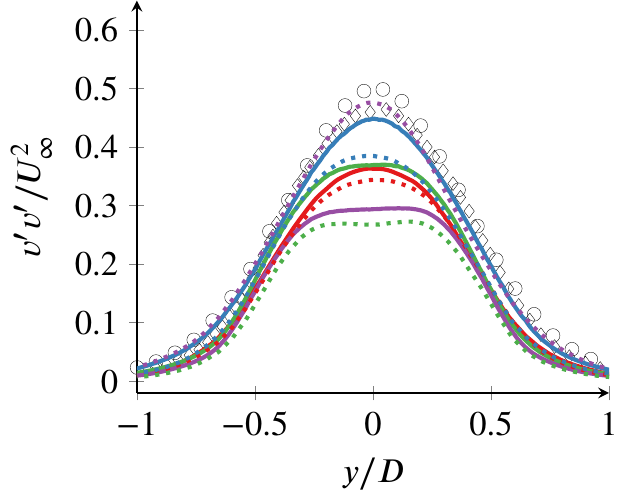}}}
        ~
        \subfloat[$x/D=2.02$]{\label{fig:uv_x2p02} \adjustbox{width=0.3\linewidth,valign=b}{\includegraphics[]{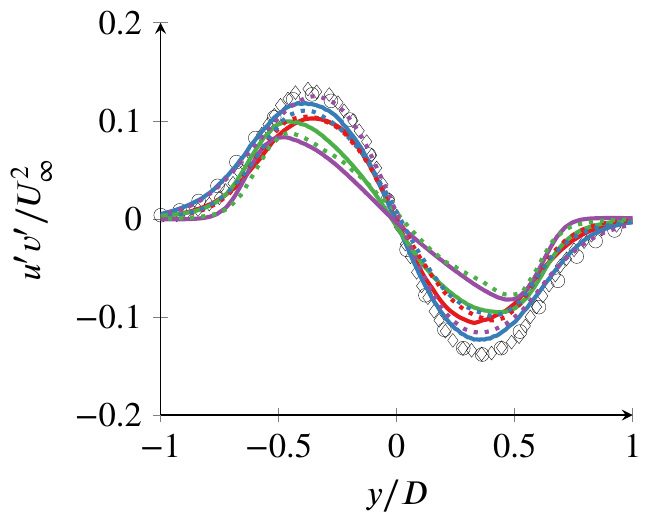}}}
        ~
        \newline
        \caption{\label{fig:cyl_fluc}Time and span-averaged streamwise velocity variance (left), normal velocity variance (middle), and streamwise-normal velocity covariance (right) profiles at $x/D = 1.06$, $1.54$, and $2.02$ using a $\mathbb{P}_3$ (solid line) and $\mathbb{P}_1$ FR scheme (dotted line).}
    \end{figure}

For the second-order statistics, the streamwise and normal velocity variance profiles as well as the streamwise-normal velocity covariance profiles at $x/D = 1.06$, $1.54$, and $2.02$ are shown in \cref{fig:cyl_fluc}. Between the various quantities and locations, the effects of the PANS model as well as the order of approximation were not consistent. Overall, although the low-order and high-order URDNS approach gave reasonable approximations of the second-order statistics at $x/D = 1.06$, the effects of the discretization as well as the introduction of the PANS model were more evident further away from the cylinder.
Except for the case of $f_k = 0.3$, the switch from a low-order to a high-order approach generally tended to increase the magnitude of the second-order statistics, likely due to the decrease in numerical dissipation. 
The accuracy of the discretization did not necessarily affect the accuracy of the second-order statistics, but instead it primarily impacted the optimal value of $f_k$, with the low-order approach showing optimal results at $f_k = 0.3$ and the high-order approach showing optimal results at $f_k = 0.1$. 

\begin{figure}[tbhp]
        \centering
        \subfloat[$\mathbb{P}_3$]{
        \adjustbox{width=0.49\linewidth,valign=b}{\includegraphics[]{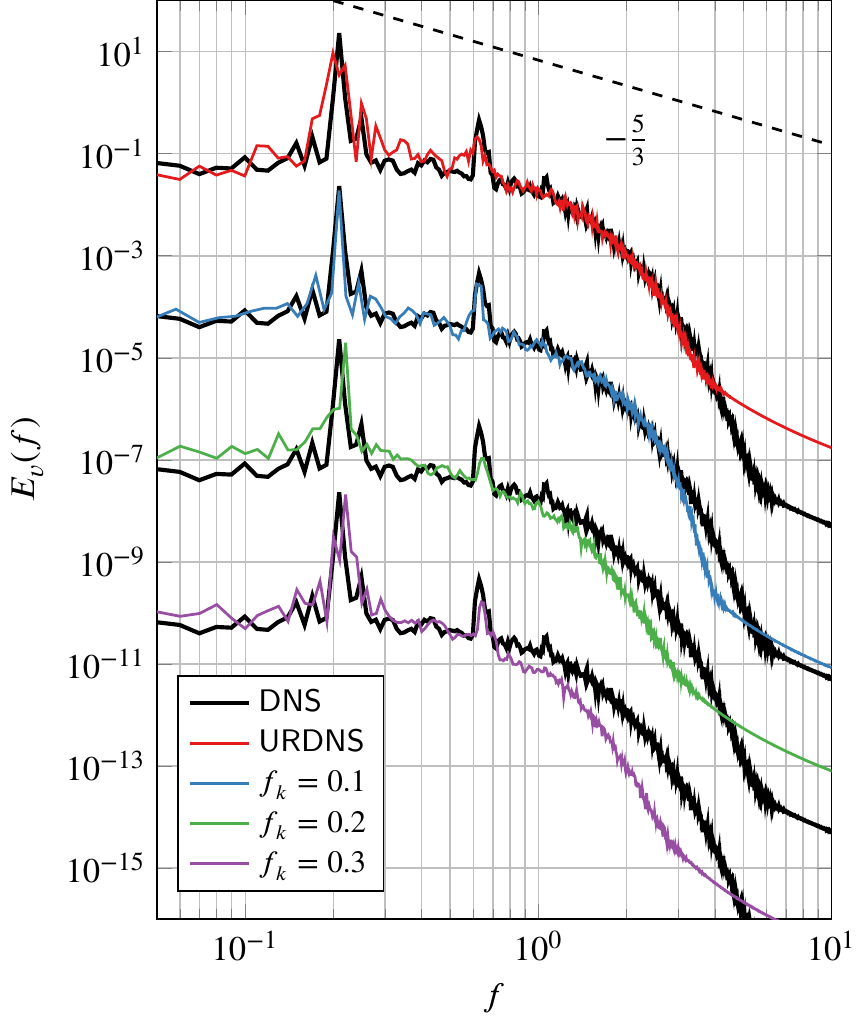}}}
        \subfloat[$\mathbb{P}_1$]{
        \adjustbox{width=0.49\linewidth,valign=b}{\includegraphics[]{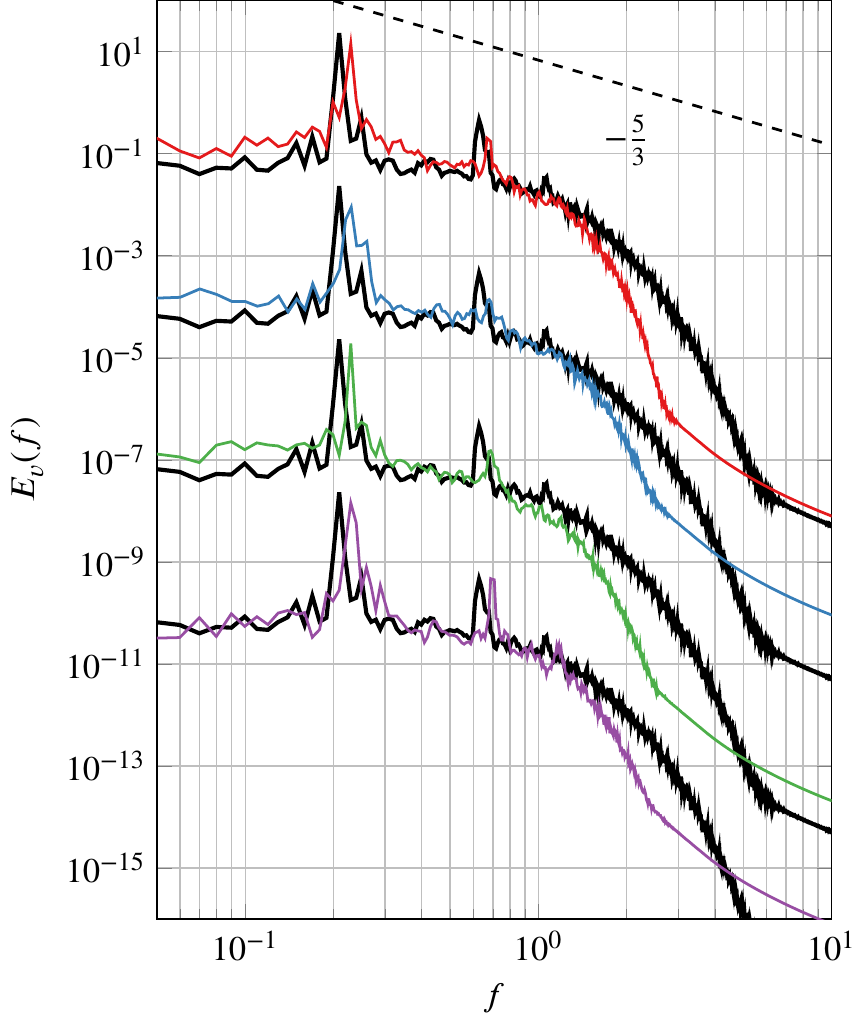}}}
        \caption{\label{fig:u_spectra_x3}Power spectra of the centerline normal velocity fluctuations at $x/D = 3$. A scaling factor of $10^{-3}$ is applied between profiles. The frequency is nondimensionalized by $D/U_\infty$. }
    \end{figure}

\begin{figure}[tbhp]
        \centering
        \subfloat[$\mathbb{P}_3$]{
        \adjustbox{width=0.49\linewidth,valign=b}{\includegraphics[]{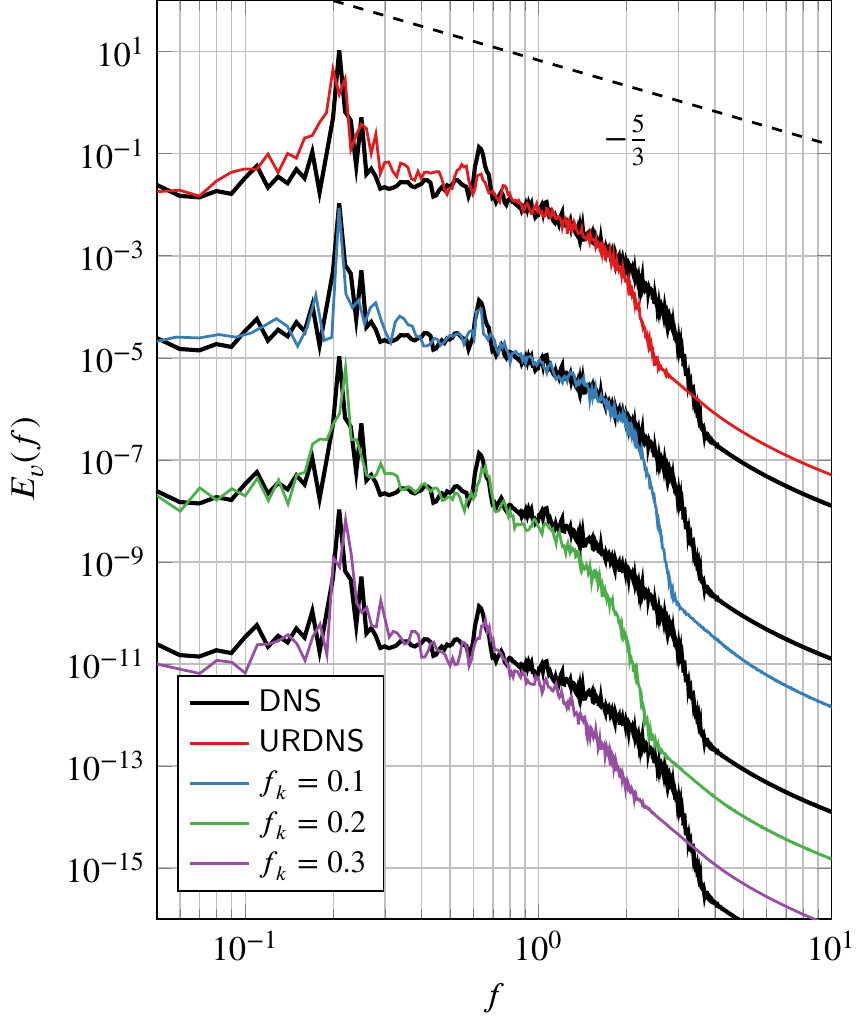}}}
        \subfloat[$\mathbb{P}_1$]{
        \adjustbox{width=0.49\linewidth,valign=b}{\includegraphics[]{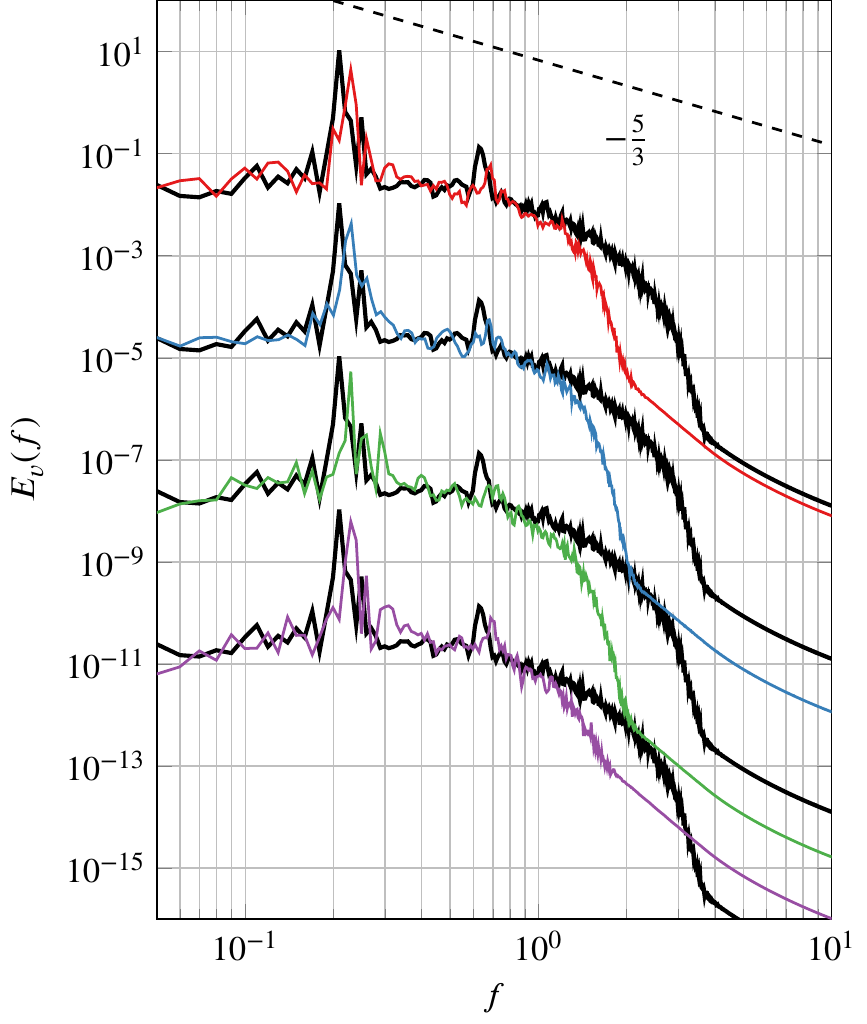}}}
        \caption{\label{fig:u_spectra_x7}Power spectra of the centerline normal velocity fluctuations at $x/D = 7$. A scaling factor of $10^{-3}$ is applied between profiles. The frequency is nondimensionalized by $D/U_\infty$.}
    \end{figure}

Due to the relative resolution of the grids, the first and second-order statistics were captured reasonably well across the various approaches used. For a more extensive comparison of these approaches, we instead focus on the flow physics and the characteristics of the coherent structures of the flow. The power spectra of the centerline normal velocity fluctuations at $x/D = 3$ and $x/D = 7$ were analyzed to evaluate the ability of the various methods to predict the frequencies of the vortex shedding and Kelvin--Helmholtz instabilities, as shown in \cref{fig:u_spectra_x3} and \cref{fig:u_spectra_x7}. From the DNS results, the Strouhal number was determined to be $St = 0.209$, which is in agreement with the numerical and experimental results of \citet{Parnaudeau2008}, and the frequency of the Kelvin--Helmholtz instability was determined to be $f_{KH} = 0.668$. At $x/D = 3$, the frequency spectra of the low-order approaches showed two distinct peaks corresponding to the vortex shedding and Kelvin--Helmholtz frequencies. However, the peak frequencies were overpredicted by roughly 10\% and 6\% for the vortex shedding and Kelvin--Helmholtz frequencies, respectively. Furthermore, except for the $f_k=0.3$ case where the peak of the Kelvin--Helmholtz instability was prominent, the low-order methods generally did not adequately resolve the frequency of the Kelvin--Helmholtz instability. For the high-order methods, the frequencies of the vortex shedding and Kelvin--Helmholtz instability were inline with the DNS results. A more substantial improvement in the PANS results was observed in comparison to the URDNS results. For all values of $f_k$, both of the peaks were significantly more prominent than in the URDNS spectra where the peaks were distributed across a larger frequency range. Additionally, there was notably less sensitivity in the spectra to the $f_k$ parameter with the high-order method than the low-order method, with $f_k = 0.1$ showing excellent agreement with the DNS results and $f_k = 0.2-0.3$ showing reasonable agreement. 

At $x/D = 7$, where the propagation of the flow over a larger computational domain presents more challenges in resolving capability, these effects were amplified. The low-order approaches again overpredicted the vortex shedding and Kelvin--Helmholtz frequencies, but at this position further away from the cylinder, the Kelvin--Helmholtz frequency was not prominently resolved by neither URDNS nor the PANS approaches. With the high-order method, an even proportionally larger improvement was observed in the PANS approaches than the URDNS approach at this position. The URDNS approach did not adequately resolve the Kelvin--Helmholtz frequency and the peak of the vortex shedding frequency was distributed across a larger frequency range. However, the accuracy of the high-order PANS approaches did not deteriorate, with both the vortex shedding and Kelvin--Helmholtz frequencies prominently resolved and the spectra for all values of $f_k$ showing good agreement with the DNS results.

    
\begin{figure}[tbhp]
        \centering
        \subfloat[DNS]{\label{fig:u_pod_dns1} \adjustbox{width=0.4\linewidth,valign=b}{\includegraphics[width=\textwidth]{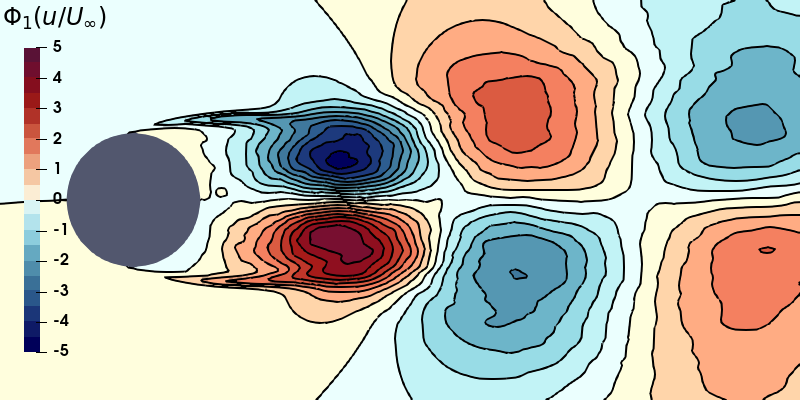}}}
        ~
        \subfloat[DNS]{\label{fig:u_pod_dns2} \adjustbox{width=0.4\linewidth,valign=b}{\includegraphics[width=\textwidth]{figs/data/cyl_POD/DNS_U1.png}}}
        ~
        \newline
        \subfloat[$\mathbb{P}_3$ URDNS]{\label{fig:u_pod_les_HO} \adjustbox{width=0.4\linewidth,valign=b}{\includegraphics[width=\textwidth]{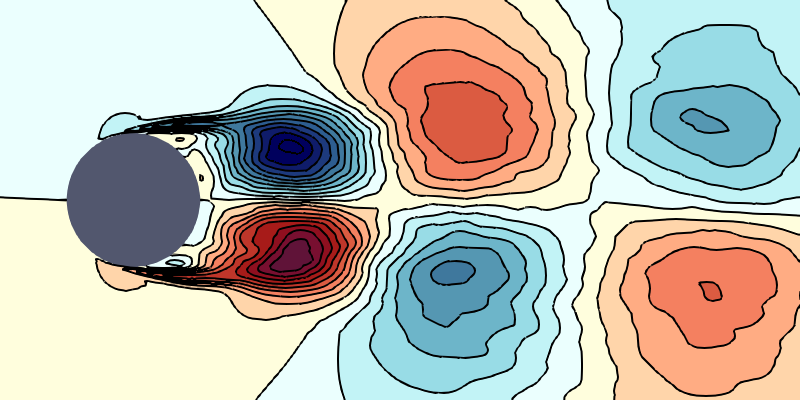}}}
        ~
        \subfloat[$\mathbb{P}_1$ URDNS]{\label{fig:u_pod_les_LO} \adjustbox{width=0.4\linewidth,valign=b}{\includegraphics[width=\textwidth]{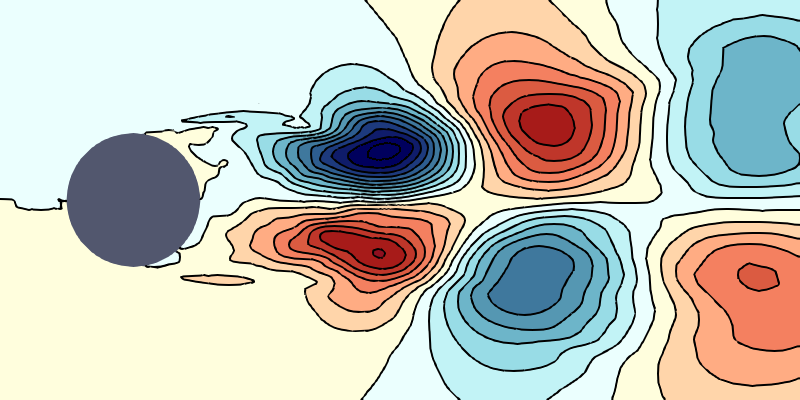}}}
        ~
        \newline
        \subfloat[$\mathbb{P}_3$ $f_k = 0.1$]{\label{fig:u_pod_fk01_HO} \adjustbox{width=0.4\linewidth,valign=b}{\includegraphics[width=\textwidth]{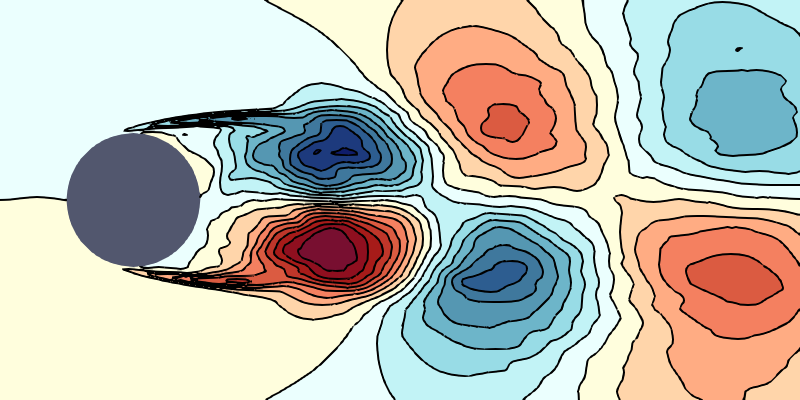}}}
        ~
        \subfloat[$\mathbb{P}_1$ $f_k = 0.1$]{\label{fig:u_pod_fk01_LO} \adjustbox{width=0.4\linewidth,valign=b}{\includegraphics[width=\textwidth]{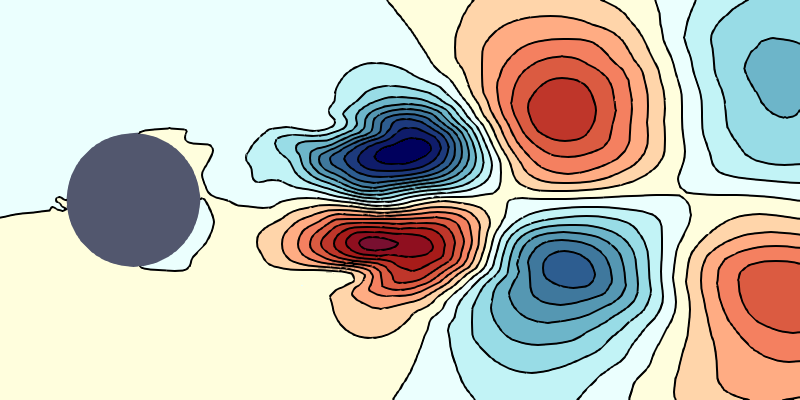}}}
        ~
        \newline
        \subfloat[$\mathbb{P}_3$ $f_k = 0.2$]{\label{fig:u_pod_fk02_HO} \adjustbox{width=0.4\linewidth,valign=b}{\includegraphics[width=\textwidth]{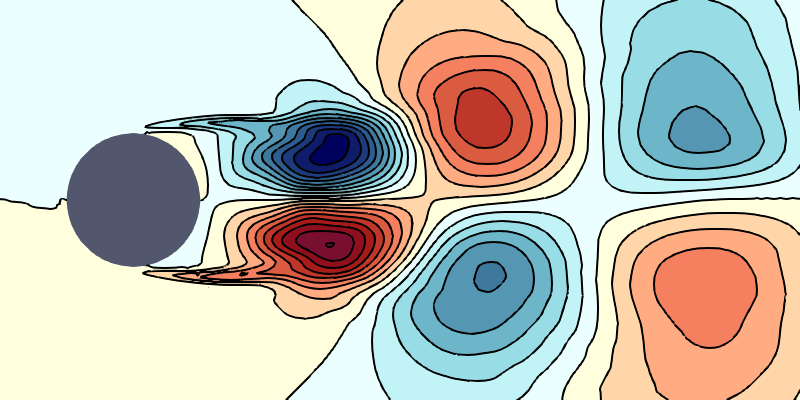}}}
        ~
        \subfloat[$\mathbb{P}_1$ $f_k = 0.2$]{\label{fig:u_pod_fk02_LO} \adjustbox{width=0.4\linewidth,valign=b}{\includegraphics[width=\textwidth]{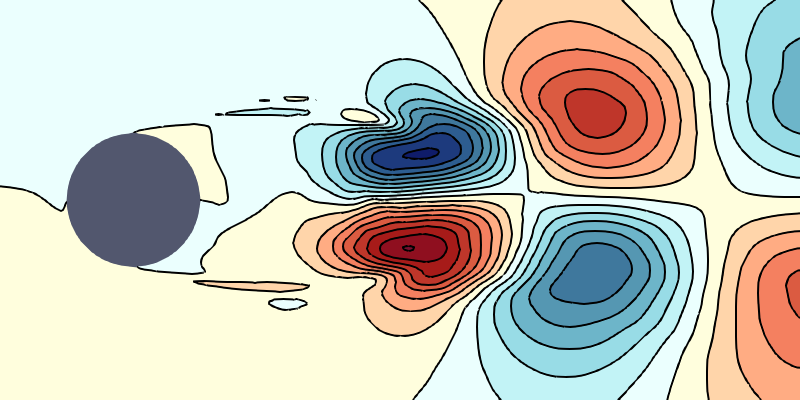}}}
        ~
        \newline
        \subfloat[$\mathbb{P}_3$ $f_k = 0.3$]{\label{fig:u_pod_fk03_HO} \adjustbox{width=0.4\linewidth,valign=b}{\includegraphics[width=\textwidth]{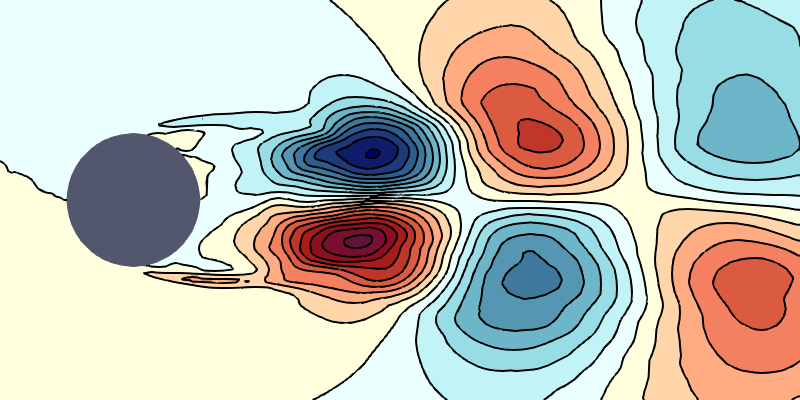}}}
        ~
        \subfloat[$\mathbb{P}_1$ $f_k = 0.3$]{\label{fig:u_pod_fk03_LO} \adjustbox{width=0.4\linewidth,valign=b}{\includegraphics[width=\textwidth]{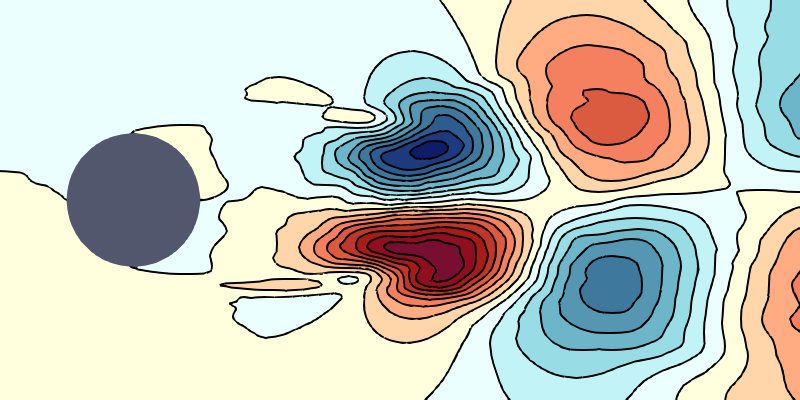}}}
        ~
        \newline
        \caption{\label{fig:cyl_uPOD}Isocontour maps of the primary POD mode of the streamwise velocity. Contour lines represent 20 equispaced subdivisions across the range. DNS results are repeated across the top row. }
    \end{figure}

For an evaluation of the ability of the various methods in predicting the coherent structures in the flow, the primary POD mode of the streamwise velocity fluctuations was compared, shown in \cref{fig:cyl_uPOD}. The primary POD mode from the DNS results depicts regions of strong correlation that are anti-symmetric across the centerline, indicative of anti-phased vortex shedding. Strong correlations in the separation region between $x/D = 2$ and $x/D = 3$ were observed as well as along the shear line emanating from the cylinder surface. Further along the wake, these correlated regions became less concentrated. Between the various approaches, the primary differences were in the positioning and shape of the strongly correlated region in the wake (i.e., the vortex shedding region) and the presence of the strongly correlated region in the shear layer. For the low-order method, the POD modes as predicted by the URDNS and PANS approaches only appreciably differed in the location of the vortex shedding region. Both approaches predicted this location farther aft of the DNS results regardless of the value of $f_k$, with the overprediction ranging monotonically from the minimum with the URDNS approach to the maximum with $f_k = 0.3$. Furthermore, neither the PANS nor the URDNS approaches replicated the circular shape of the vortex shedding region effectively, and the presence of a strongly correlated region in the shear layer was not seen. With the high-order method, these issues were generally rectified, with all approaches showing reasonable agreement with the DNS results in terms of the shape of the vortex shedding region and the presence of the strongly correlated region in the shear layer. As with the low-order method, the differences between the URDNS and PANS approaches were primarily with respect to the location of the vortex shedding region. The URDNS approach predicted this region much closer to the cylinder in comparison to the DNS results, whereas the PANS approaches generally showed good agreement with the DNS results. Due to the low variation in the results between various $f_k$ values, no significant decrease in sensitivity to the $f_k$ parameter was observed with the high-order approach. 

\section{Conclusion}\label{sec:conclusion}
In this work, the effects of high-order discretizations on a hybrid turbulence model were explored in the context of scale-resolving simulations of turbulent flows. The PANS approach was discretized using the FR scheme and employed on two canonical benchmarks: the wall-bounded, separated flow around a periodic hill and the wake flow around a circular cylinder at $Re = 3900$. By varying the order of the approximation while fixing the total degrees of freedom, the effects of the discretization error on the PANS approach was independently investigated and compared to under-resolved DNS approaches. 

In general, the switch from a low-order to a high-order approximation tended to proportionally improve both the URDNS and PANS approaches equally with respect to the first-order statistics. However, for highly under-resolved simulations such as the periodic hill in the present work,
the high-order discretization improved the PANS prediction of the second-order statistics notably more than the URDNS prediction. For the more complex wake flow problem around the cylinder, the grid resolution was proportionally higher, and therefore the focus of the comparison was placed on the prediction of the flow physics since the first and second-order statistics could be reasonably approximated regardless of the methods used. The prediction of the flow physics improved significantly more through a high-order approximation using PANS than with URDNS, with much better prediction of the frequencies of the vortex shedding and Kelvin-Helmholtz instabilities, especially at distances farther along the wake. Furthermore, high-order approximations of the PANS approach showed a larger improvement in the prediction of the dominant POD mode of the streamwise velocity than high-order approximations of the URDNS approach.

Overall, high-order approximations tended to benefit the PANS approach proportionally more than the URDNS approach, as larger improvements in the prediction of the statistics and flow physics were generally seen with PANS. These benefits were attributed to the lower numerical dissipation of the high-order schemes, which allowed for better resolution of the small-scale features predicted by the model equations that can be dissipated by low-order schemes. 
Additionally, less sensitivity to the resolution-control parameter was observed with the high-order PANS approach, resulting in less variation in the predictions than with the low-order approximation. These findings indicate that high-order discretizations may be an effective approach for increasing the accuracy and reliability of hybrid turbulence models for scale-resolving simulations without a significant increase in computational effort.

\section*{Acknowledgements}
\label{sec:ack}
This research did not receive any specific grant from funding agencies in the public, commercial, or not-for-profit sectors.

\bibliographystyle{unsrtnat}
\bibliography{reference}


\clearpage
\begin{appendices}
\crefalias{section}{appendix}
\section{Turbulence Model Parameters}\label{app:constants}
The constants for the SST model are as tabulated by \citet{Menter1994}. Certain constants in the SST model are defined with an inner (1) and outer (2) constant, such that its value is defined as 
\begin{equation*}
    \phi = F_1 \phi_1 + (1 - F_1)\phi_2
\end{equation*}
for the following inner and outer values
\begin{alignat*}{4}
    &\alpha_1 &&= 0.5532, &&\quad\quad  \alpha_2 &&=0.4403,\\
    &\beta_1 &&= 0.075,  &&\quad\quad \beta_2 &&=0.0828,\\
    &\sigma_{k1} &&= 0.85, &&\quad\quad \sigma_{k2} &&=1.0,\\
    &\sigma_{\omega1} &&= 0.5, &&\quad\quad  \sigma_{\omega2} &&=0.856.
\end{alignat*}
The remaining constants are explicitly defined as 
\begin{equation*}
    a_1 = 0.31, \quad \quad \beta^* = 0.09, \quad \quad Pr_t = 0.9.
\end{equation*}
\section{Proper Orthogonal Decomposition}\label{app:pod}

The POD method, as described by \citet{Weiss2019}, can be used to decompose a time-dependent velocity fluctuation field, $u'(\mathbf{x}, t)$, into a set of orthonormal spatial modes, $\Phi_k(\mathbf{x})$, such that 
\begin{equation}
    u'(\mathbf{x}, t) = \sum_{k=1}^\infty a_k(t) \Phi_k(\mathbf{x}),
\end{equation}
for some temporal coefficient $a_k(t)$. In the discrete form, this is performed by first forming a matrix of "snapshots" of the solution,
\begin{equation}
    \mathbf{U} = 
\begin{pmatrix}
    u_{11} & \hdots & u_{1n}\\
    u_{21} & \hdots & u_{2n}\\
    \vdots & \ddots & \vdots\\
    u_{m1} & \hdots & u_{mn}
\end{pmatrix}
= 
\begin{pmatrix}
    u'(\mathbf{x}_1, t_1) & \hdots & u'(\mathbf{x}_n, t_1)\\
    u'(\mathbf{x}_1, t_2) & \hdots & u'(\mathbf{x}_n, t_2)\\
    \vdots & \ddots & \vdots\\
    u'(\mathbf{x}_1, t_m) & \hdots & u'(\mathbf{x}_n, t_m)\\
\end{pmatrix},
\end{equation}
where $n$ denotes the number of sample points in space and $m$ denotes the number of sample points in time. A correlation matrix $\mathbf{C}$ can then be defined as
\begin{equation}
    \mathbf{C} = \frac{1}{m-1}\mathbf{U}\mathbf{U}^T.
\end{equation}
By performing an eigendecomposition of $\mathbf{C}$, a set of eigenvectors $\mathbf{\Psi}_k$ and their associated eigenvalues $\lambda_k$ can be extracted and sorted such that $|\lambda_k| > |\lambda_{k+1}|$. The spatial modes are then defined as 
\begin{equation}
    \mathbf{\Phi}_k = \mathbf{U}^T \mathbf{\Psi}_k
\end{equation}
and normalized such that $\|\mathbf{\Phi}_k\|_2 = 1$.

In this work, the snapshots were formed by resampling the flow field to a two-dimensional grid on the spanwise periodic boundary. 800 and 200 points were sampled in the streamwise and normal directions, respectively, and 20,000 snapshots were obtained over the time-averaging horizon, corresponding to values of $n = 160,000$ and $m = 20,000$. 

\end{appendices}


\end{document}